\documentclass[aps,pre,preprint,showpacs,amssymb,showpacs,preprintnumbers,amsmath,superscriptaddress]{revtex4}

\newcommand{\myfigscale}{0.3}

\usepackage[dvips]{epsfig,graphicx}
\usepackage{amsmath,amssymb,amsfonts,wasysym,color}
\usepackage{subfigure}

\begin{document}
\title{Effect of 
  ions on confined near-critical binary aqueous mixtures}
\author{Faezeh Pousaneh}
\affiliation{Institute of Physical Chemistry, Polish Academy of Sciences,
  Kasprzaka 44/52, PL-01-224 Warsaw, Poland}
\author{Alina Ciach}
\affiliation{Institute of Physical Chemistry, Polish Academy of Sciences,
  Kasprzaka 44/52, PL-01-224 Warsaw, Poland}
\author{Anna Macio{\l}ek}
\affiliation{Max-Planck-Institut f{\"u}r Intelligente Systeme,
  Heisenbergstra{\ss}e 3, 70569 Stuttgart, Germany}
\affiliation{Universit{\"a}t Stuttgart, Institut f{\"u}r Theoretische und Angewandte Physik,
  Pfaffenwaldring 57, 70569 Stuttgart, Germany}
\affiliation{Institute of Physical Chemistry, Polish Academy of Sciences,
  Kasprzaka 44/52, PL-01-224 Warsaw, Poland}
\begin{abstract} 

 Near-critical  binary mixtures 
containing ions and confined between two charged and selective surfaces are studied within a Landau-Ginzburg theory 
extended to include 
electrostatic interactions. Charge density profiles and the effective interactions between the confining 
surfaces are  calculated in the case of chemical preference of ions for  one 
 of the solvent components. Close 
to the consolute point of the binary solvent, the preferential solubility of ions leads to the modification of the charge
 density profiles in respect to the ones obtained from the Debye-H\"uckel theory.
 As a result, the electrostatic contribution to the effective potential between the charged surface 
can exhibit an attractive well. Our calculations are based on  the approximation scheme 
 valid if the bulk correlation length of a  solvent is much larger than the Debye screening length; 
in this critical regime  the effect of charge on the concentration profiles of the solvent is subdominant.
Such conditions are met in the recent measurements of
  the effective forces acting between a substrate and a spherical colloidal particle immersed in the near-critical
 water-lutidine mixture
[Nature {\bf 451}, 172 (2008)].
 Our analytical results are in a quantitative agreement with the experimental ones.
\end{abstract}

\maketitle
\section{Introduction}

Ions dissolved in a binary liquid mixture often display  preferential  solubility  in the  one  component  of the  solvent.
Also the interaction between the $\pm$ ions and the solvent can be different, which is known as an unequal partitioning of ions.
In a bulk system, a selective solvation leads to the shift of the critical point of the
demixing transition
 \cite{onuki:04:01,eckfeldt:43:0,hales:66:0} and to a number of other effects on the phase separation of a binary solvent
\cite{onuki:11:00}. 
In the presence of   external charged surfaces
the shift of the bulk critical point of a mixture can be  enhanced by a dielectric inhomogeneity 
arising due to the attraction of
high permittivity solvent  to the charged surface (dielectrophoretic forces)
 \cite{tsori:07:01,podgornik:09:01,reviews}. 
Moreover,  a selective solvation of ions may change the  concentration profiles of the binary solvent near the wall and, reversely,
  adsorption phenomena can significantly influence the distribution   of ions near a charged surface.
These mutual influences have been recently studied theoretically for the case when the binary solvent is near its consolute point 
in the semi-infinite geometry \cite{cm:2010},  and for systems confined between two parallel walls or substrates \cite{pc:2011,bgod:2011,st:2011}.
In the latter case, the consequences of the ions-solvent coupling for the effective forces acting on the  confining
surfaces  were studied.

One of the motivation for such investigations 
is provided by recent 
experimental works  \cite{nature,nature_long,nellen}, where the effective potential between a
 charged substrate and a likely charged colloidal particle immersed in a water-lutidine critical mixture ($T_c\simeq 307.15K$) 
 was  directly measured. The surfaces of the colloidal particle and of the flat substrate with similar or opposite adsorption
 preferences were used in order to verify predictions of the theory for the thermodynamic Casimir force. 
These, so called, critical Casimir forces acting between the colloidal particle and a flat substrate
arise as a result of the modifications of the relevant
 order parameter (OP) and restrictions of its 
fluctuation spectrum by the confining surfaces.
Close to the critical point of the  solvent, attraction is predicted for like surfaces, whereas  repulsion
 is predicted if one surface is  hydrophilic and the other one is
hydrophobic. 

The theory of  effective interactions between two surfaces confining a near-critical fluid 
is well  developed for uncharged surfaces and
 for mixtures of neutral components  \cite{FdG,krech:99:0,Dbook,gambassi}. 
However, in the  experiments mentioned above the surfaces were charged, and moreover,
a small amount of 
ions was present in the solution. In Ref.~\cite{nature,nature_long} the ions result from dissociation of water, 
and in Ref.~\cite{nellen} a  hydrophilic salt was added.  Far from $T_c$  repulsion has  been present
independently of the adsorption preferences of the surfaces, because the electrostatic potential
 dominates \cite{nature,nature_long,nellen}. The electrostatic repulsion decays exponentially with the decay rate 
equal to the Debye screening length $1/\kappa$.
 For $T\to T_c$, in addition to the repulsion for small separations
 $L\sim 1/\kappa$,   an
attraction (repulsion) has been observed for surfaces with the same (opposite) 
adsorption preferences for larger separations, $L\sim \xi$, where $\xi$ is the bulk correlation length of the solvent
\cite{nature,nature_long}.
 Such behavior is predicted by
 the sum of the electrostatic and the critical Casimir potentials for the corresponding boundary conditions. 
A full quantitative agreement between the experiment
 and the sum of the electrostatic and the critical Casimir potentials
 could not be obtained, however \cite{nature,nature_long}. 
The sum of the electrostatic potential that fitted well the experimental results far from $T_c$  and of 
the critical Casimir potential that fitted well the data for separations $L\gg 1/\kappa$ close to $T_c$, 
for intermediate distances disagreed strongly with the measured potential.
The authors concluded that coupling between the critical 
concentration  fluctuations  and the 
distribution of ions may lead to modifications of the potential. For this reason, only  distances
significantly larger than the screening length were considered close
to the critical point  to verify the theory of the critical Casimir potential. 
In the presence of salt a more complex behavior was observed, in particular, an unexpected
 attraction between  hydrophilic and  hydrophobic surfaces for intermediate temperatures ($\xi\kappa<1$) \cite{nellen,pc:2011,bgod:2011,st:2011}. 

The shape of the effective interaction potentials between charged selective surfaces confining the critical binary solvent 
with ions resembles strongly the intermolecular interaction potentials, but on a much larger scale.
 Because of possible applications, the ability to design the interaction potential between, e.g., the two colloidal particles is of interest,
 therefore the mutual effect of ion distribution and concentration profiles deserves serious attention.

Theoretical  studies reported in Refs.~\cite{pc:2011,bgod:2011,st:2011,st:2012}  are all
 based on the Ginzburg-Landau-like theory but with a different level of complexity as far as  the 
parameter space is concerned. In general, a high-dimensional parameter space is 
required for a full description of a four-component mixture, with two of the species carrying a charge,  in a presence
of two charged   and selective surfaces.
In  Refs.~\cite{bgod:2011,st:2011} the  reduced description has been  employed  in order to investigate the particular mechanisms and the role of the specific interactions.
Accordingly, in these studies, e.g., 
the van der Waals (vdW) type of interactions  between ions and between ions and the walls have been neglected altogether.
 In  Ref.~\cite{bgod:2011},
a non-trivial interplay between critical and electrostatic phenomena (which goes beyond the simple superposition of the
critical Casimir and the electrostatic potentials)
 arises as a result of an unequal partitioning of the salt ions in a non-uniform
solvent. In Ref.~\cite{st:2011}, the focus is on the electrostatic effects, therefore also
 interactions between the components of the solvent and the walls have been neglected.
A preference of charged walls for one of  the solvent components (with the largest permittivity) 
has been  taken into account via the composition-dependent 
permittivity. Within this approach, for an equal  partitioning
of the salt ions in  each component of the  solvent, an attraction between like-charge surfaces can occur
as a result of  dielectrophoretic  forces and the ion-solvent coupling.

Here we extend the theoretical  approach developed in Ref.~\cite{cm:2010} for a semi-infinite system 
to the {\it slit} geometry  and  determine the influence of critical adsorption on the charge distribution close to 
the critical point of the solvent, i.e., for 
$\kappa\xi>1$.
Such a ratio of  relevant
 length scales in the system  has been realized in the  experiments  described in Ref.~\cite{nature,nature_long}. 
Next we examine the effect of these modifications {of the distribution of ions} on the form of the effective 
potential between 
confining surfaces which are charged and selective. Within  the approximation scheme that we use in our analysis, this  effective potential 
 can be written as a sum of three contributions: the critical Casimir potential, the pure electrostatic potential (as given by the linearized Debye-H\"uckel (DH)
theory), and the potential arising from the ion-solvent coupling.  We use a Derjaguin approximation  \cite{derjaguin}  in order to compare  our theoretical predictions for the effective potential 
with the experimental data reported in 
Ref.~\cite{nature,nature_long}. Within the Derjaguin approximation the interaction potential between the sphere and the planar wall
is expressed in terms of the interaction potential in the slit. For the critical Casimir part of the total effective potential we use
the scaling function  determined to a great degree of accuracy from
 the MC   simulations  in $d=3$ \cite{vasiliev}.

The  description of the system used in the present paper is more  complete than the ones used in Refs.~\cite{bgod:2011,st:2011,st:2012} 
 in the sense that  it
treats ions as the molecules which interact also non-electrostatically with each other and with the walls.
Consequently, we consider a system confined by two charged walls which are selective to {\it all} components of the mixture.
Due to the more  complete description, the  parameter space of our model is  somewhat larger than in the other approaches  
 \cite{bgod:2011,st:2011,st:2012}.  
 In the full version of the model~\cite{cm:2010}, the
vdW interactions between all pairs of components of the mixture, and the dependence of the permittivity on 
the concentration was assumed. The number of parameters can be reduced for particular systems.
 For example, for hydrophilic ions we are left with 3 parameters characterizing non-Coulombic interactions~\cite{cm:2010},
 while two such parameters are present in Refs.  \cite{bgod:2011,st:2011}.   
Rather general description developed from microscopic theory lends itself to  still another  mechanisms 
leading to the  unintuitive effects in the slit geometry \cite{pc:2011}.  
Moreover, the general framework of our theory is also suitable for antagonistic salt,  which leads to interesting phenomena~\cite{sonks}.
Finally, because the  Ginzburg-Landau-type theory that we employ has been developed from the microscopic lattice gas model of the 
four-component mixture, in our model the entropy of mixing is better approximated than in Refs.~\cite{bgod:2011,st:2011}, 
where the entropy of mixing has been 
taken separately for the binary solvent (without ions)  and separately for the ions (as an entropy of an  ideal gas).
In the present work we assume, as in Ref.~\cite{bgod:2011} a uniform permittivity, because the dielectrophoretic effects can be mimicked by an appropriate 
contributions  to the surface fields. Here we consider the case of the equal partitioning of the ions
 in the solvent (like in Ref.~\cite{st:2011}).
What distinguishes our study from  the other similar approaches proposed recently,
is that our {\it analytical} results are  obtained {\it beyond} the linear approximation for the EL equations, and a {\it quantitative}, 
not only a qualitative agreement with experiments is obtained.   
 As in Ref.\cite{st:2012}, interesting effects appear when the nonlinear terms in the EL equations are included.

Our presentation is organized as follows.
 In  Sec.~\ref{sec:background}, we provide  the physical background of the  phenomena studied in the present work.
In Sec.~\ref{sec:GL} we describe our model. Approximate, analytical solutions of the Euler-Lagrange equations for the order parameters,
 valid for $\kappa\xi>1$,  are given and discussed in Sec.~\ref{sec:sol}. In Sec.~\ref{sec:pot} we obtain results for the effective potential
between the confining surfaces. The quantitative comparison with the experimental data are described in Sec.~\ref{sec:exp}.
We discuss our results and conclude in Sec.~\ref{sec:con}.

\section{Background}
\label{sec:background}

A wall of a container or a surface of a colloidal particle disturb the structure of the fluid in contact with them because 
of geometrical constraints 
on the positions of the molecules,  and because the molecules interact with the matter 
of the wall rather than with the fluid molecules that are missing beyond the external surface. Structural changes 
are present for separations from the surface of the order of the bulk correlation length $\xi$. In particular, 
near a surface preferentially 
adsorbing one component of the mixture the excess concentration of this component extends to distances $\sim\xi$, and for 
$T\to T_c$ (hence $\xi\to\infty$) this phenomenon is called critical adsorption~\cite{diehl:86:0}.

When a second, parallel  wall at the separation $L$ from the first one is present, the  excess grand potential of the fluid 
confined in the slit 
has the form~\cite{evans:90:0}
\begin{eqnarray}
\label{Omega}
\Omega_{ex}=\omega_{ex}A=\Omega+pAL=((\gamma_0+\gamma_L)+\Psi(L))A
\end{eqnarray}
where $A$ is the area of each surface,  $p$ is the bulk pressure and $\gamma_0$, $\gamma_L$  are the surface tensions at the 
corresponding walls. The surface tension results
 from the particle-wall interactions, and from
 the modification  of the structure of the fluid
near the single surface in the semiinfinite system. The effective potential $\Psi(L)$ reflects the mutual effect of both 
surfaces on
 the structure of the fluid. The structure of the fluid  is influenced simultaneously
 by both  walls  if $L\sim \xi$, therefore $\Psi(L)$ vanishes
 for $L\gg\xi$. Since the confined fluid tends to minimize the grand potential, $\Psi(L)$  and $-\nabla \Psi(L)$ act
 as an effective 
potential and an effective force between the confining surfaces respectively.
Close to the critical point associated with either gas-liquid or demixing transition of the confined fluid $\xi\to\infty$ and  
$\Psi(L)$ acquires a  universal contribution  which becomes long-ranged at the critical point. 
This contribution to  $\Psi(L)$ is termed the critical  Casimir potential and  exhibits scaling
described  by a universal  scaling function  which
is determined solely by  the so-called universality class  of the
phase transition occurring in the
bulk, the geometry, and the surface 
universality classes of the confining surfaces. The range of $\Psi(L)$ can be tuned by small temperature changes, because
$\xi=\xi_0\tau^{-\nu}$, where
$\tau=(T-T_c)/T_c$,  $T_c$ is the critical temperature of the solvent, 
 the  critical exponent is $\nu\approx 0.63$  and the system-dependent parameter $\xi_0$ is
of order of a few Angstroms. 
 The theory of the thermodynamic Casimir force \cite{FdG,krech:99:0,Dbook,gambassi}, based on 
the theory of critical phenomena
in confinement \cite{FSS,diehl:86:0} is well established. 
For the Ising universality class in a slit  geometry, pertinent to the present study,   $\Psi(L)$  
decays exponentially with the decay 
length equal to the bulk correlation length for  distances $L\gtrsim \xi$.  
For the symmetrical (antisymmetrical) surfaces the potential is attractive (repulsive). 
In the case of a binary mixture,  symmetrical (antisymmetrical) surfaces have the same (opposite) adsorption preferences
 for the components of the binary mixture. Following a  convention used commonly in the literature 
we denote by $(+,+)$  (and, equivalently, $(-,-)$) the boundary conditions (BC) 
which reflect the fact that the two  surfaces effectively attract the same component of a liquid mixture, whereas
 $(+,-)$ BC correspond to the case in which the two   surfaces  attract  different  components.

According to the above discussion, one expects the effective critical Casimir interaction  to occur
between  a  colloidal particle and  a  planar wall  or  between two colloidal particles immersed 
 in a near-critical binary solvent. 
Often, in such systems also electrostatic interactions are present, e.g., 
in colloidal suspensions that are  charge-stabilized.
 The charge at the  colloidal particles or at the charged wall
is screened by the counterions in the solvent. Accordingly,  the electrostatic interactions  between 
two charged colloidal particles or between a colloidal particle and a charged wall 
 become   exponential functions of the distance and can  compete with  the critical Casimir forces.
For instance, the effective interaction between charged  planar  surfaces  decays as
$\pm\exp(-\kappa L)$, where the repulsion (attraction) corresponds to the likely (oppositely) charged surfaces, 
and the dimensionless inverse Debye screening length 
is
\begin{equation}
 \label{kappa}
 \kappa^* =a \kappa =\sqrt{\frac{4\pi e^2\rho_c^*}{k_BT\bar\epsilon}}.
\end{equation}
$\rho_c^*=\rho_c a^3$ is the dimensionless number density of ions and  $a$ is the microscopic length unit
(we shall choose for $a$ the size of the solvent molecules). 
Moreover, this competition can become an interplay.
In the present paper we consider  charged surfaces immersed in a  binary solvent.
In such systems, if  the solubility of ions in both components of the mixture is the same, then the distribution of charges is independent 
of the local solvent 
concentration, and also the concentration of the mixture is not affected by the presence of the ions. As a consequence, 
the presence
of charges at the confining surfaces has no effect on the critical Casimir potential  and, vice versa, 
 the critical adsorption has no effect on the 
electrostatic interactions between the charged surfaces. Therefore $\Psi(L)$ is just a sum of the critical Casimir 
and the electrostatic potentials. Usual salts, however, are soluble in water and
 insoluble in organic liquids.  In such a case the critical adsorption of the component preferred by the wall and the
 distribution 
of charges may influence each other, and as a result may lead to a different form of $\Psi(L)$. The room-temperature critical
 points are present in  mixtures of water and organic liquids, therefore the question how the critical adsorption and the
distribution of hydrophilic ions influence each other and modify $\Psi(L)$ is of practical importance.

\section{Ginzburg-Landau theory} 
\label{sec:GL}

In this section, following Ref.~\cite{cm:2010} we briefly summarize the main steps in developing   the Ginzburg-Landau theory both
from  a microscopic 
lattice gas model and from  a continuum one.

\subsection{Derivation of the model}
\label{subsec:der}

In order to obtain the Ginzburg-Landau functional from a continuum microscopic model, one starts
from the grand thermodynamic potential of the four-component mixture \cite{evans:90:0}
\begin{eqnarray}
\label{OmegaL}
\Omega
=U_{SR}+U_{el}-TS-\int_{V}d{\bf r}\mu_i\rho_i({\bf r}),
\end{eqnarray}
 where $U_{SR}$  is the  energy associated with the short-range (SR)   vdW
 interactions, 
 $U_{el}$ is the  electrostatic energy, $S$ is the entropy,  $T$ is the  temperature and $\mu_i$  is the chemical potential of 
the $i $-th species.  Local dimensionless number densities  are denoted by $\rho^*_i({\bf r})=\rho_i({\bf r})a^3$, 
where $i=1,2,3,4$ for water, oil, $+$ and $-$ for ions, respectively. For $a$ we have chosen the diameter of the organic molecules. 
In equilibrium, $\rho^*_i({\bf r})$
 correspond to the minimum of $\Omega$ for given $T$, $\mu_i$ and the boundary conditions. 
For ionic species of the same valence, $\mu_3=\mu_4=\mu_c$ because of the charge-neutrality condition. 
Integration (summation in the lattice version)  in Eq.~(\ref{OmegaL}) is over the system volume $V=AL$, and summation 
convention for
 repeated indices is assumed in the whole paper.
We assume the usual form of the internal energy $U_{SR}$,
\begin{eqnarray}
\label{USR}
U_{SR}
=Au_{SR}=\int_{V}d{\bf r}\int_{V} d{\bf r}' \frac{1}{2}\rho^*_i({\bf r})V_{ij}({\bf r}-{\bf r}') g_{ij}({\bf r}-{\bf r}')
\rho^*_j({\bf r}')\\
\nonumber
 +\int_{V}d{\bf r}\rho^*_i({\bf r})V_{i}^s({\bf r}),
\end{eqnarray}
where $V_{ij}$ and $g_{ij}$ are the vdW
interaction and the pair correlation function between the corresponding components respectively,
 and $V_{i}^s({\bf r})$ is the sum of the direct wall-fluid potentials acting on the
 component $i $.  In our model length is in $a$ units, i.e. we consider  dimensionless $r^*=r/a$ in (\ref{USR}) and in the whole article. However, to simplify the notation we drop the asterisk for $r$ as well as for the characteristic lengths 
(like $\kappa^{-1}$, see (\ref{kappa})). It should be remembered that  length is dimensionless.
 In the lattice model only nearest-neighbors interact, and the integration in Eq.~(\ref{USR})  should be replaced by a
 summation.
In continuum we assume that both the zeroth and the second moments, $ V_0^{(ij)}=\int d{\bf r}g_{ij}(r) V_{ij}(r)$ and 
 $ V_2^{(ij)}=\frac{1}{6}\int d{\bf r}g_{ij}(r)V_{ij}( r)r^2$, respectively,  are finite. 

Compressibility of the liquid can be neglected, so  we assume $\sum_{i=1}^4\rho_i^*=1$. 
The three independent  densities can be chosen as: a concentration of the solvent,
\begin{eqnarray}
\label{s}
 s=\rho_1^*-\rho_2^*, 
\end{eqnarray}
a dimensionless density of the solute,
 \begin{eqnarray}
\label{rhocdef}
  \rho_c=\rho_3^*+\rho_4^*
 \end{eqnarray}
(subscript $c$ from 'charge') and a dimensionless charge density,
 \begin{eqnarray}
\label{phidef}
\phi=\rho_3^*-\rho_4^*.
  \end{eqnarray}

Based on the experimental case where ions in the solution come from dissociation of a water, a similar chemical nature  
of the anion and the cation is assumed in Ref.~\cite{cm:2010}, and  any difference between the interactions of 
the anion or the cation and any other species is neglected. In the case of salts insoluble in organic liquids the 
above assumption is not strictly valid, and should be considered as an approximation, whose validity should be verified 
at the later stage. 
This assumption distinguishes our analysis from  Ref.~\cite{bgod:2011}, and has an important consequence for the form of 
the short-range interaction energy. Namely,  $U_{SR}$ expressed in terms of the new variables
depends only on $s$ and $\rho_c$, and is
 independent of $\phi$~\cite{cm:2010}, as can be verified by assuming  $V_{i,3}= V_{i,4}$  in Eq.~(\ref{USR}).

The electrostatic energy in a slit with the surface charge $\sigma(n)$  
at the $n$-th  wall ($n=0,L$) is 
\begin{eqnarray}
\label{DH}
\frac{ U_{el}}{A}=u_{el}=\int_0^{L}dz \left[ -\frac{\epsilon}{8\pi}(\bigtriangledown\psi)^2
+e\phi  \psi \right]\\
\nonumber
+e\sigma(0)\psi(0)+
e\sigma(L) \psi(L),
 \end{eqnarray} 
where $e$ is the elementary charge, $\epsilon$ is the dielectric constant of the solvent and the electrostatic potential 
 $\psi$ satisfies the Poisson equation,
\begin{eqnarray}
\label{Poisson}
 \frac{\epsilon}{4\pi}\frac{d^2 \psi(z)}{d z^2}+e\phi(z)= 0.
\end{eqnarray}
We neglect the dependence of $\epsilon$ on the solvent 
concentration for two reasons. Firstly, 
 we take into account that in the critical region the amplitude of the 
deviations from the average concentration is small an hence such a dependence 
leads to the higher order  corrections to the order parameter profiles, which  we neglect (see Ref.~\cite{cm:2010} ).
Secondly, as already mentioned in the Introduction,  the dielectrophoretic effects can be mimicked by an appropriate
contributions  to the surface fields.  Accordingly, while comparing our results with the experimental data we treat the surface fields as the
fitting parameters.
 For an analysis of the dielectrophoretic effects
 see Refs.~\cite{bap,podgornik:09:01,st:2011,reviews}.

The entropy $S$ in the lattice model has the form of the ideal mixing entropy. Here we assume the same approximation.

\subsection{Separation of the charge-dependent and charge-independent parts of the grand potential}
\label{subsec:sep}

The theory of critical phenomena was developed for uncharged  systems, 
therefore we shall separate the part depending on the charge density from the remaining part of the grand potential. 
For the latter part we shall apply the Ginzburg-Landau description.

In the new variables  (Eqs.~(\ref{rhocdef}) and (\ref{phidef}))
$S$ can be split 
into two terms,
\begin{eqnarray}
 S=-k_B A\sum_{i=1}^4\int_0^L dz\rho_i^*(z)\ln\rho_i^*(z)= (s_C[s,\rho_c]+s_{el}[\rho_c,\phi])A,
\end{eqnarray}
with
\begin{eqnarray}
\label{sC}
s_C[s,\rho_c]= -k_B \int_0^L dz\Bigg[ \frac{1-\rho _c(z)+s(z)}{2}\ln\left(\frac{1-\rho_c(z)+s(z)}{2} \right)\\\nonumber
 +\frac{1-\rho _c(z)-s(z)}{2}\ln\left(\frac{1-\rho_c(z)-s(z)}{2}\right)\\\nonumber
+ \rho _c(z)\ln\Big(\frac{\rho _c(z)}{2}\Big)\Bigg ]
\end{eqnarray}
and
 \begin{eqnarray}
\label{sel}
s_{el}[\rho_c,\phi]=- k_B \int_0^L dz\Bigg[ 
 \frac{\rho _c(z)+\phi(z)}{2}\ln\left(\frac{\rho _c(z)+\phi(z)}{2} \right)\\\nonumber
 +\frac{\rho _c(z)-\phi(z)}{2}\ln\left(\frac{\rho _c(z)-\phi(z)}{2}\right)-\rho _c(z)\ln\Big(\frac{\rho _c(z)}{2}\Big)\Bigg ].
\end{eqnarray}
We use the subscript $'el'$ for the  quantities that are directly 
or indirectly associated with electrostatics and vanish for
 $\phi=0$, and  the subscript $'C'$ is from ``Casimir''.
From the above properties it follows that the grand potential is a sum of the two terms
\begin{eqnarray}
\label{ocoel}
\Omega= (\omega_C[s,\rho_c] +\omega_{el}[\rho_c,\phi])A
 \end{eqnarray}
where
\begin{eqnarray}
\label{oel}
 \omega_{el}[\rho_c,\phi]=u_{el}[\phi]-Ts_{el}[\rho_c,\phi]
\end{eqnarray}
and
\begin{eqnarray}
\label{OMC}
 \omega_C[s,\rho_c] = u_{SR}[s,\rho_c] -Ts_C[s,\rho_c] -\int_0^L dz \mu_i\rho_i(z) 
\end{eqnarray}
Note that $s_C[s,\rho_c]+k_B\ln 2 \int_0^L dz \rho_c(z)$ (see Eq.~(\ref{sC}))  equals the entropy density of a three-component 
charge-neutral mixture with the solute density $\rho_c$ (i.e., $\rho_3^*=\rho_4^*=\rho_c/2$)  and
 the solvent concentration $s$ in the case of close packing. Using this observation,  one can interpret 
  Eq.~(\ref{OMC}) as  the grand-potential density of such a three-component neutral mixture 
with $\mu_c $ replaced by $ \mu_c+k_BT\ln 2$  (recall that $\mu_c=\mu_3=\mu_4$). 
Thus, we have separated  from the grand potential $\Omega$ the contribution independent of the charge, which has this advantage that the 
 methods developed for neutral near-critical systems can be directly applied to $\omega_C[s,\rho_c] $.

Let us focus on Eq.~(\ref{oel}), which can be rewritten as
\begin{eqnarray}
 \omega_{el}[\rho_c,\phi]=f_{el}[\rho_c,\phi]-f_{el}[\rho_c,0]
\end{eqnarray}
where
\begin{eqnarray}
  f_{el}[\rho_c,\phi]=u_{el}[\phi]+k_BT \int_0^L dz\Bigg[ 
(1-\rho _c(z))\ln(1-\rho_c(z)) \\\nonumber
 +\frac{\rho _c(z)+\phi(z)}{2}\ln\Big(\frac{\rho _c(z)+\phi(z)}{2} \Big)
 +\frac{\rho _c(z)-\phi(z)}{2}\ln\Big(\frac{\rho _c(z)-\phi(z)}{2}\Big )\Bigg].
\end{eqnarray}
 This term alone describes the ions dissolved in a {\it homogeneous} solvent of the  density $1-\rho_c$.
However, the two
contributions in Eq.~(\ref{ocoel}), $\omega_C$ and $\omega_{el}$,  are coupled through $\rho_c$. 
In equilibrium,  $\rho_c(z)$ corresponds to the minimum of $\Omega$ (Eq.~(\ref{ocoel})), 
therefore  the critical  Casimir and the electrostatic contributions to the effective potential are not independent, 
even in the case of identical chemical nature of the ions.  
This implies that the effective potential between the confining walls {\it must}
 differ from the sum of the critical Casimir potential and of  the electrostatic potential in the case of  a homogeneous solvent.

\subsection{Expansion of the functional}
\label{sec:expan}

 For given $T$ and $\mu_i$ the bulk equilibrium densities, $\bar s$, $\bar \rho_c$ and $\bar \phi=0$, 
 correspond to the minimum of the bulk part of $\Omega$, $-pAL$.
Deviations of the fields 
$s$ and  $\rho_c$ from the bulk equilibrium values  are denoted by 
\begin{eqnarray}
\label{excessdens}
 \vartheta_1(z)=s(z)-\bar s\\
\vartheta_2(z)= \rho_c(z)-\bar \rho_c
\end{eqnarray}
where $z$ is the distance from the left wall. 
In equilibrium $\vartheta_1(z)$, $\vartheta_2(z)$ and $\phi(z)$ correspond to
 the minimum of $\omega_{ex}$  (see Eq.~(\ref{Omega})). Close to the 
 critical point  $\vartheta_1(z)$, $\vartheta_2(z)$ and $\phi(z)$ are small for $z\sim\xi$, 
therefore the entropy can be Taylor expanded 
and the expansion can be truncated. From Eq.~(\ref{sel}) we have for fixed $\bar\rho_c$

\begin{eqnarray}
s_{el}[\bar\rho_c+\vartheta_2,\phi]=s_{DH}[\phi]+\Delta s[\vartheta_2,\phi]
\end{eqnarray}
where
\begin{eqnarray}
s_{DH}[\phi]=- k_B \int_0^L dz\Bigg[\sum_{n\ge 1} a_{n}\phi^{2n}(z)
 \Bigg ]
\end{eqnarray}
and 
\begin{eqnarray}
\Delta s[\vartheta_2,\phi]=- k_B \int_0^L dz\Bigg[ \sum_{n\ge 1}\sum_{m\ge 1} a_{n,m}\phi^{2n}(z)\vartheta_2^m(z)
 \Bigg ].
\end{eqnarray} 
The above form follows from the fact that we have chosen to split the total entropy in such a way that  $s_{el}[\rho_c,\phi]$ vanishes for $\phi=0$.
The coefficients $a_n$ and $a_{n,m}$ resulting from the Taylor expansion of Eq.~(\ref{sel}) are functions of $\bar\rho_c$.
The above equations  for fixed $\bar\rho_c$ yield
\begin{eqnarray}
\label{Lex}
 {\cal L}_{el}[\vartheta_2,\phi]=\omega_{el}[\bar\rho_c+\vartheta_2,\phi]=
{\cal L}_{DH}[\phi]+\Delta {\cal L}[\vartheta_2,\phi]
\end{eqnarray}
 where 
\begin{eqnarray}
\label{LDH}
 {\cal L}_{DH}[\phi]=u_{el}[\phi]+k_B T\int_0^L dz\Big[\frac{\phi^2 }{2\bar \rho_c}+O(\phi^4)\Big]
\end{eqnarray}
 and
\begin{eqnarray}
\label{DeltaL}
 \Delta {\cal L}[\vartheta_2,\phi]=-T\Delta s[\vartheta_2,\phi]=
-k_B T\int_0^{L}dz \Bigg[\frac{\vartheta_2(z)\phi^2(z)}{2\bar \rho^2_c}+O(\vartheta_2^2\phi^2,\vartheta_2\phi^4)\Bigg].
\end{eqnarray}
As already mentioned, Eqs.~(\ref{Lex})-(\ref{DeltaL}) with (\ref{DH}) and (\ref{Poisson}) describe the ionic 
system with the charge density 
$\phi(z)$ 
and the total density of ions $\bar\rho_c+\vartheta_2(z)$, placed between parallel charged walls.
 When the second term in Eq.~(\ref{Lex})
 is neglected,  no excess of the number density of ions at the surfaces is obtained. It is 
 the term (\ref{DeltaL}) of purely entropic origin that leads
to the excess number density of ions near the surfaces~\cite{israelachvili:10:0,barrat:03:0} when the chemical nature 
(and hence the interactions with the wall) of the anion and the cation are the same. 
%This term   leads in turn to the mutual effect of the excess solvent concentration and the charge profile in the case of the near-critical solvent.

From the above considerations it follows that the excess grand potential can be split into three terms,
\begin{eqnarray}
\label{omexap}
 \omega_{ex}[\vartheta_1,\vartheta_2,\phi]\approx {\cal L}_C[\vartheta_1,\vartheta_2] +{\cal L}_{DH}[\phi]
+\Delta {\cal L}[\vartheta_2,\phi],
\end{eqnarray}
where
$ {\cal L}_C[\vartheta_1,\vartheta_2]=\omega_C[\bar\rho_c+\vartheta_2,\bar s+\vartheta_1]
-\omega_C[\bar\rho_c,\bar s]$. Since $ {\cal L}_C$ describes
 the near-critical
two-component solvent with addition of one kind of neutral solute, it can be approximated by the
 Landau-type functional
by using standard coarse-graining procedures. 
Close to the critical temperature  $\vartheta_1(z)$ and $\vartheta_2(z)$ vary on the length
 scale  large compared to the molecular size, and 
 $\vartheta_1(z'),\vartheta_2(z')$  can be Taylor expanded about $z'=z$. 
From Eqs.~(\ref{USR}) and (\ref{sC}) we thus  obtain \cite{cm:2010}
\begin{eqnarray}
\label{LC}
 {\cal L}_C={\cal L}_C^0+k_BT\int_0^{L}dz \sum_{n}\sum_{m}b_{n,m} \vartheta_1^{2n}(z)\vartheta_2^m(z),
\end{eqnarray}
where in the summation in Eq.~(\ref{LC}) $2n+m\ge 3$, $b_{n,m}$ are functions of $\bar\rho_c$, and 
\begin{eqnarray}
\label{LC0}
 {\cal L}_C^0[\vartheta_1,\vartheta_2]=\frac{1}{2}\int_0^{L}dz\Bigg\{\vartheta_i(z)C^0_{ij}\vartheta_j(z)
+\nabla\vartheta_i(z) J_{ij}\nabla\vartheta_j(z)
\Bigg\} \\
\nonumber +\frac{\vartheta_i(0)J_{ij}\vartheta_j(0)}{2}-\bar h_i(0)\vartheta_i(0)
%\\\nonumber
+\frac{\vartheta_i(L)J_{ij}\vartheta_j(L)}{2}-\bar h_i(L)\vartheta_i(L),
\end{eqnarray}
 where
\begin{eqnarray}
\label{C0}
 C^0_{ij}=- J^0_{ij}-T\frac{\partial^2 s_C}{\partial \vartheta_i\partial \vartheta_j}_{|\vartheta_i=0,\vartheta_j=0},
\end{eqnarray}
 $J_{ij}^0=\int d{\bf r} J_{ij}( r)$ and $J_{ij}=\frac{1}{6}\int d{\bf r} J_{ij}( r)r^2$. $- J_{ij}(r)$ represents the vdW
 interactions for $\vartheta_i$ and $\vartheta_j$, and can be obtained from the vdW contribution to Eq.~(\ref{OmegaL}) with
 the densities expressed in terms of the new variables
(see (\ref{s}) -(\ref{phidef})). 
 We assume the same interaction ranges for all interacting  pairs and postulate
 $J_{ij}^0=6J_{ij}$ (recall that we consider dimensionless distance).
 Explicit expressions of $ C^0_{ij}$ are given in
 Ref.~\cite{cm:2010} and in  Appendix A.  
 Finally, 
\begin{eqnarray}
\label{hbar}
\bar h_i(n)= h_i(n)- J_{ij}\vartheta_j(n),
\end{eqnarray}
 where 
$ h_i(n)$ are the surface fields describing direct interactions with the $n$-th wall. $J_{ij}\vartheta_j(n)$
 and the remaining surface terms in Eq.~(\ref{LC0})  compensate for the interactions with the missing fluid
neighbors due to the presence of  the wall;  such  interactions  are present in the bulk term, but  should be replaced by the interactions with
 the molecules of the wall   \cite{cm:2010}.

 When the mixture phase separates,  both the solvent concentration $s$ and the density of the solute $\rho_c$ are different in
 the coexisting phases, because of a much bigger solubility of  the solute  in   water. Likewise, for $T$ close to 
$ T_c$ both $s$ and $\rho_c$ exhibit long-range  critical fluctuations. 
Thus, in the  Fourier representation the bulk part of ${\cal L}_C^0$ can be written in the form
\begin{eqnarray}
\label{LC0eigen}
A{\cal L}_C^0[\vartheta_1,\vartheta_2]\equiv AL_C^0[\Phi_1,\Phi_2]=\int d{\bf k} \frac{1}{2}\Big[\tilde \Phi_1(-{\bf k})\tilde C_1(k)\tilde \Phi_1({\bf k})
+\tilde \Phi_2(-{\bf k})\tilde C_2(k)\tilde \Phi_2({\bf k})\Big]
\end{eqnarray}
where $\tilde C_i(k)$ and $\tilde \Phi_i({\bf k})$ are the eigenvalue and the eigenvector of $\tilde C_{ij}(k)=C^0_{ij}+k^2J_{ij}$, respectively.
  The critical order parameter, $\tilde \Phi_1({\bf k})$,
 is associated with the eigenvalue $\tilde C_{1}(k)$  that vanishes  at $T_c$; $\tilde C_{2}(k)$  at $T_c$ is positive and of the
order of unity. The asymptotic decay of correlations in a real space is dominated by $\tilde C_{1}(0)\propto \xi^{-2}$.
In the critical region,  the  contribution from the noncritical OP  $\tilde \Phi_2({\bf k})$ 
to the grand potential is much larger than the contribution 
from the critical OP. Accordingly,  the probability of  fluctuations associated with $\Phi_2$
 is negligible compared to the probability of the fluctuations corresponding to the critical OP. This  allows  us to
neglect the noncritical fluctuations and    
 $L_C^0[\Phi_1,0]$ takes  the usual form associated  with the critical Casimir potential for  the Ising universality class.
 From the computational point of view, in the present work  it is more convenient to consider both fields, $\vartheta_1$ and $\vartheta_2$,
 instead of their linear combination $\Phi_1$. 

\section{Approximate solutions of the Euler-Lagrange equations}
\label{sec:sol}

In this section we derive the approximate EL equations for the functional (\ref{omexap}) and obtain approximate 
solutions for the solvent concentration, the solute density and the charge in a slit of width $L\sim\xi$. In the one-phase 
region we neglect the second term in (\ref{LC}), and consider the lowest-order approximation which incorporates  the coupling between
the critical adsorption and the distribution of charges,
\begin{eqnarray}
\label{omexap0}
 \omega_{ex}[\vartheta_1,\vartheta_2,\phi]\approx {\cal L}_C^0[\vartheta_1,\vartheta_2] +{\cal L}_{DH}[\phi]+\Delta {\cal L}[\vartheta_2,\phi].
\end{eqnarray}
The first, second and third terms on the  RHS  of Eq.~(\ref{omexap0}) are given by  Eqs.~(\ref{LC0}), (\ref{LDH}) and (\ref{DeltaL}), respectively. 
We neglect the higher order terms in Eqs.~(\ref{LDH}) and (\ref{DeltaL}), and in this approximation the expansion of the 
functional is truncated at the third order term in the fields $\vartheta_i$ and $\phi$.

The Euler-Lagrange equations, obtained by minimization
of the approximate functional (\ref{omexap0}) with respect to the  fields $\vartheta_i$ and $\phi$, together with the
 Poisson equation (\ref{Poisson}), take a  rather simple form \cite{pc:2011},
\begin{eqnarray}
\label{EL1}
 \frac{d^2 \vartheta_i(z)}{d z^2}=M_{ij}\vartheta_j(z) +d_i\phi^2(z) 
\end{eqnarray}
\begin{eqnarray}
\label{EL2}
\frac{d^2 \phi(z)}{d z^2}=\kappa^2\phi(z) +\frac{1}{\bar\rho_c}\frac{d^2 (\phi(z)\vartheta_2(z))}{d z^2}.
\end{eqnarray}
In the above $M_{ij}=(J^{-1})_{ik}C^0_{kj}$, where $(J^{-1})_{ik}$ is the $(i,k)$-th element of the matrix
 inverse to the matrix $J_{ij}$~\cite{cm:2010}, and
$(d_1,d_2)=-\frac{k_BT}{2\bar\rho_c^2}\Big( (J^{-1})_{12},( J^{-1})_{22}\Big)$.
 The solutions must satisfy the charge neutrality condition, 
\begin{eqnarray}
\label{chargeneut}
\int_0^Ldz \phi(z)+\sigma_0+\sigma_L=0,
\end{eqnarray}
  and the boundary conditions   for $\vartheta_i$: \cite{cm:2010}
\begin{eqnarray}
\label{bc0}
\frac{d \vartheta_i(z)}{d z}|_{z=0}-\vartheta_i(0)= H_i(0)
\\
\nonumber
-\frac{d \vartheta_i(z)}{d z}|_{z=L}-\vartheta_i(L)=H_i(L),
\end{eqnarray}
where 
\begin{eqnarray}
\label{Hi}
H_i(n)=-(J^{-1})_{ij}\bar h_j(n),
\end{eqnarray}
and $\bar h(n)$ is defined in Eq.(\ref{hbar}). For a hydrophilic (hydrophobic) wall $H_1<0$  ($H_1>0$). 
Consistently with the approximate form of the functional (\ref{omexap0}), the RHS of 
Eqs.~(\ref{EL1}) and (\ref{EL2}) are truncated at the second order terms.

\subsection {\bf  Solutions of the linearized EL equations }
The linearized equations (\ref{EL1}) and (\ref{EL2}) for $\vartheta_i$ and $\phi$ are decoupled.  
The solutions take the well known forms
\begin{eqnarray}
\label{philin}
\phi^{(1)}(z)=-\frac{\kappa\sigma_0}{1-e^{-\kappa L}}\Big(e^{-\kappa z}+R_\sigma e^{-\kappa(L-z)}\Big),
\end{eqnarray} 
where we  denote the ratio of the surface charges at the two surfaces by 
\begin{equation}
\label{eq:rsigma} 
R_{\sigma}=\frac{\sigma_{ L}}{\sigma _{0}} 
\end{equation}
and use the superscript $(1)$  to distinguish the solutions of the linearized equations.
The above charge profile 
obeys the charge neutrality condition (\ref{chargeneut}). The corresponding approximation for the electrostatic potential is
(see linearized Eq.~(\ref{EL2}), and Eqs.~(\ref{Poisson}), (\ref{kappa}))
\begin{eqnarray}
\label{psi1}
 \psi^{(1)}(z)=-\frac{kT}{\bar\rho_c}\phi^{(1)}(z).
\end{eqnarray}

The  excess concentration of the solvent  and the excess number density of ions (\ref{excessdens}) in the critical region 
$T \to T_c $ ($\xi \to \infty$)  take the approximate form
\begin{eqnarray}
\label{vlin}
 \vartheta_1^{(1)}=t_0 e^{- z/\xi}+t_L e^{- (L-z)/\xi},\\\nonumber
\vartheta_2^{(1)}=n_0 e^{- z/\xi}+n_L e^{-(L-z)/\xi}.
\end{eqnarray}
Because  in the critical region the  decay length $\lambda^{-1}$ associated with the larger eigenvalue 
$\tilde C_{2}(0)\propto\lambda^2$ is negligible 
compared to $\xi$, the  terms  $\propto e^{-\lambda z}$ are subdominant and can 
 be omitted for slits with $L\gg a$. 
From the boundary conditions  we obtain the approximate expressions
\begin{eqnarray}
\label{n0}
n_{0}\simeq n_{01}\frac{1-C R_ne^{-L/\xi}}{1-C^2 e^{-2L/\xi}}\\\nonumber
n_{L}\simeq   n_{01}\frac{R_n- C e^{-L/\xi}}{1-C^2 e^{-2L/\xi}}\\\nonumber
t_{0}\simeq t_{01}\frac{ 1-C R_te^{-L/\xi}}{1-C^2 e^{-2L/\xi}}\\\nonumber
t_{L}\simeq  t_{01}\frac{R_t- C e^{-L/\xi}}{1-C^2 e^{-2L/\xi}}
\end{eqnarray}
where
\begin{eqnarray}
C=\frac{(\xi-1)}{(\xi+1)}
\end{eqnarray}
 and 
\begin{eqnarray}
\label{rn}
n_{01}=\frac{-H_{2}(0)\xi}{(\xi+1)}\simeq_{\xi\to \infty } -H_{2}(0),\hskip1cm t_{01}\simeq_{\xi\to \infty } -H_{1}(0),
\\\nonumber
R_{n}=\frac{H_{2}(L)}{H_{2}(0)}, \quad R_{t}=\frac{H_{1}(L)}{H_{1}(0)}.
\end{eqnarray}
We note that the decoupling of the fields  $\vartheta_2$ and $\phi$ that occurs after linearization of  EL equations is rather
unphysical. Nonlinear terms are necessary in order to  regain the right physics.

\subsection {\bf Leading-order corrections in the critical region}

In this section we  determine the leading-order corrections to the solutions of the linearized 
 EL equations (\ref{philin}) and (\ref{vlin}).
In Ref.~\cite{cm:2010} it was assumed that except from distances  $\sim a$ from each wall the dimensionless fields
 $f=\vartheta_i,\phi$ are  all of the same order of magnitude, $f=O(\nu)$, where $\nu$ is a small parameter. 
$\vartheta_i$ and $\phi$ are proportional to $H_{i}$ and $\sigma$ respectively, thus the analysis in Ref.~\cite{cm:2010} 
 is restricted to the surfaces with $H_{i},\sigma=O(\nu)$. Under the above assumption   analytical solution of the EL 
equations can be obtained by systematic approximations within
a perturbation method. Since the RHS of Eqs.~(\ref{EL1}) and (\ref{EL2}) are truncated according to the truncation of 
the functional $\omega_{ex}$ (see (\ref{omexap0})), in a consistent approximation the solutions
should have the form $\vartheta_i=\vartheta_i^{(1)}+\vartheta_i^{(2)}$ and $\phi_i=\phi_i^{(1)}+\phi_i^{(2)}$. 
The superscript $(2)$ refers to the leading order correction terms (of order $O(\nu^2)$), which satisfy the linear
 inhomogeneous
 equations
\begin{eqnarray}
\label{EL1a}
 \frac{d^2 \vartheta^{(2)}_i(z)}{d z^2}=M_{ij}\vartheta^{(2)}_j(z) +d_i\left(\phi^{(1)}(z)\right)^2 
\end{eqnarray}
\begin{eqnarray}
\label{EL2a}
\frac{d^2 \phi^{(2)}(z)}{d z^2}=\kappa^2\phi^{(2)}(z) +\frac{1}{\bar\rho_c}\frac{d^2 (\phi^{(1)}(z)\vartheta_2^{(1)}(z))}{d z^2}.
\end{eqnarray}
In the above $\phi^{(1)}$ and $\vartheta_i^{(1)}$ are given by  Eqs.~(\ref{philin}) and (\ref{vlin}) respectively. 
The boundary conditions are $d\vartheta_i^{(2)}(z)/dz|_{z=0}=\vartheta_i^{(2)}(0)$,
 $d\vartheta_i^{(2)}(z)/dz|_{z=L}=-\vartheta_i^{(2)}(L)$ and  $\int_0^L dz\phi^{(2)}(z)=0$, because
 $\vartheta_i^{(1)}$ obey Eqs.~(\ref{bc0}), and  $\phi^{(1)}$ obeys the charge neutrality condition (\ref{chargeneut}). 
 Note that because the Poisson equation (\ref{Poisson}) is linear, from the above and (\ref{kappa}) 
 we obtain the leading-order correction to the electrostatic potential
\begin{equation}
\label{psi2}
 \psi^{(2)}=\frac{k_BT}{e\bar\rho_c}\Bigg[ \frac{\vartheta_2^{(1)}\phi^{(1)}}{\bar\rho_c}-\phi^{(2)}\Bigg].
\end{equation}
Note that  in a semiinfinite system  the second term on the RHS in Eq.~(\ref{EL1a}) decays as $\sim\exp(-2\kappa z)$, 
and the second term on the RHS in Eq.~(\ref{EL2a}) decays as $\sim\exp(-\kappa z)\exp(-z/\xi)$.
Further approximations are possible when one of the two length scales, either the correlation $\xi$ or the screening 
length $1/\kappa$, is much larger than the other length. Following Refs.~\cite{cm:2010,pc:2011}
 we introduce  the ratio between the correlation and the screening
 lengths,
\begin{eqnarray}
\label{y}
 y=\kappa\xi,
\end{eqnarray}
and focus on the two limiting cases: (i)  $y\ll 1$,  i.e., the Debye length is much larger than the correlation length,
 and (ii) $y\gg 1$, i.e.,  the Debye length is much smaller than the correlation length. 
The analysis of the limiting cases can be done with a reasonable effort.
 We should note that the experiments 
 showing unusual attractive effective potential between the charged colloidal particle and the  charged wall 
having  the opposite adsorption preferences,  were performed for
$y<1$ \cite{nellen}, whereas the experiments reported in Ref.~\cite{nature,nature_long} concern the case $y>1$. 
 
Let us first focus on the case (i), which was  studied in Ref.~\cite{pc:2011}.
  For $y\ll 1$, from Eq.~(\ref{vlin}) 
 we have $\vartheta_i(1/\kappa)\sim \exp(-1/y)\ll 1$, and the second term on the RHS of Eq.~(\ref{EL2}) 
can be neglected. As a result we obtain  that $\phi\approx\phi^{(1)}$, and $\vartheta_i$ satisfy Eqs.~(\ref{EL1}) and (\ref{bc0}).
The solution of  Eq.~(\ref{EL1}) with $\phi\approx\phi^{(1)}$ 
yields  a qualitative agreement with the experimental
 results for the effective potential obtained for a system with the Debye length 
 larger than the correlation length \cite{nellen}.

The case (ii) was studied in Ref.~\cite{cm:2010} for  a semiinfinite system.
 For $y\gg 1$ in the semiinfinite system we have (see Eq.~(\ref{philin})) $\left(\phi^{(1)}(\xi)\right)^2\sim\exp(-2y)\ll 1$.
 Thus
 for $z\sim\xi$ the second term on the RHS of
 Eqs.~(\ref{EL1}) and (\ref{EL1a}) is  subdominant with respect to the second term on the RHS of Eq.~(\ref{EL2a}) 
decaying as  $\sim\exp(-\kappa z)\exp(-z/\xi)\sim_{|z\sim \xi} \exp(-y-1)$. Therefore, in the asymptotic region of $y\gg 1$ we neglect the former but keep the latter.
This means that we can  approximate the concentration and the number density of ions by the solutions of the linearized EL
 equations  $\vartheta_i\approx \vartheta_i^{(1)}$, but we cannot do it for  the charge density $\phi$. 
In physical terms the effect of the charge profile on the critical
 adsorption is negligible for $y\gg 1$, 
because the neutralizing charge in the fluid is present at the distances from the surface $z\sim\kappa^{-1}\ll \xi$;  
the charge
 distribution can be neglected on the same footing as the distribution of  molecules at the distance $\lambda^{-1}$ from 
the wall.

\begin{figure}
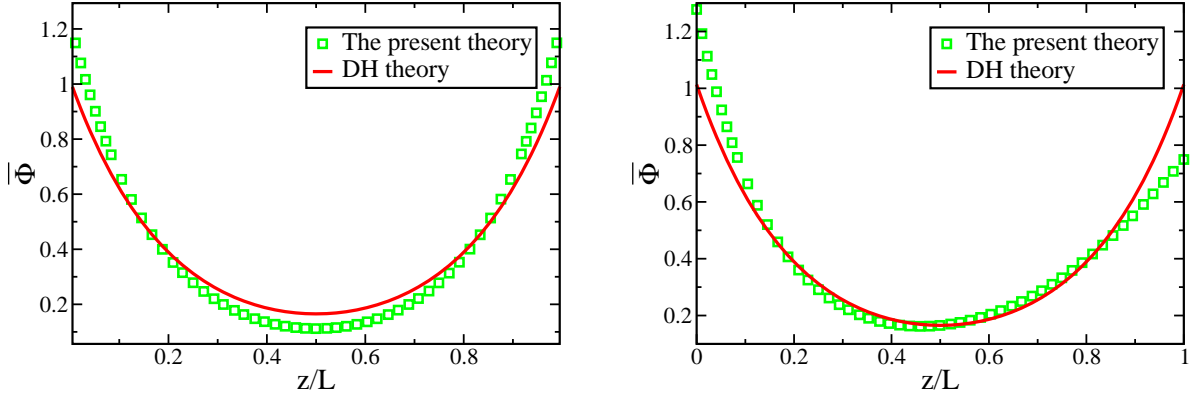

\includegraphics[scale=\myfigscale]{Fig1-a.eps} \qquad
\includegraphics[scale=\myfigscale]{Fig1-b.eps}
\caption{Charge density profile  as a function of the scaled distance $z/L$. The solid line represents $\phi^{(1)}(z)$ 
(Eq.~(\ref{philin}),
 linearized DH theory result) and the dotted line is the approximate charge density, Eq.~(\ref{phiapprox}), 
with the effect of the critical adsorption included. $\kappa \xi=5$, $\kappa L=5$ , $R_{\sigma}=|R_n|=1$ (the same charge 
densities and the same or opposite adsorption preferences at both surfaces), and the excess number density of ions at the
 surface is 
 $n_{01}=0.5$. Charge density is in $(-\kappa \sigma_0)$ units. Panels (a) and (b) correspond to the boundary conditions
  $(-,-)$ and  $(-,+)$ respectively.}
\label{fig:1}
\end{figure}

In this work we focus on the case of $y>1$, and adopt the approximation valid in the asymptotic 
region $y\gg 1$. We neglect the effect of the charge distribution on $\vartheta_1$ and $ \vartheta_2$, 
and obtain the leading-order correction to the charge profile from the approximate equation (\ref{EL2a}).
The solution of  Eq.~(\ref{EL2a}) can be written in the form 
\begin{eqnarray}
\label{eq:slit14}
  \phi^{(2)}(z) &=& \frac{\kappa\sigma_0 }{\bar\rho_c(1-e^{-\kappa
 L})}\Bigg[{{\cal A}}_0e^{-\kappa z}+{{\cal A}}_Le^{-\kappa(L- z)} 
 +{\cal A}_{00}e^{-\kappa z}(1-e^{- z/\xi})
+ {\cal A}_{0L}e^{-\kappa z}(1-e^{-(L-z)/\xi}) \nonumber  \\ 
& + & {\cal A}_{LL}e^{-\kappa\left(L- z\right)}(1-e^{-(L-z)/\xi})
 + {\cal A}_{L0}e^{-\kappa \left(L- z\right)}(1-e^{- z/\xi})\Bigg].
\end{eqnarray}
The coefficients are functions of $y$, $n_0$, $n_L$ and $R_{\sigma}$, and their explicit forms  are given in Appendix B.
The obtained approximation for the charge density
\begin{eqnarray}
\label{phiapprox}
\phi(z)\approx \phi^{(1)}(z)+\phi^{(2)}(z)
\end{eqnarray}
is shown in Figs.~\ref{fig:1}(a) and (b)   for $(-,-)$ and $(-,+)$ boundary conditions, respectively; results correspond to  $y=\kappa L=5$.

%%%%%%%%%%%%%%%%%%%%%%%%%%%%%%%%%%%%%%%%%%%%%%%%%%%%%%%%%%%%%%%%%%%%%%%%%%%%%%%%%%%%%%%%%%%%%%%%%%%%%%%%%%%%%%%%%%%%%%%%%%%%%%%%%%%%%%%%%%%%%%%%5

%%%%%%%%%%%%%%%%%%%%%%%%%%%%%%%%%%%%%%%%%%%%%%%%%%%%%%%%%%%%%%%%%%%%%%%%%%%%%%%%%%%%%%%%%%%%%%%%%%%%%%%%%%%%%%%%%%%%%%%%%%%%%%
\section {\bf The effective potential}
\label{sec:pot}

The main goal of this work is a determination of the effective potential $\Psi(L)$ between the  surfaces that are both selective 
and charged.
From Eq.~(\ref{omexap0}) it immediately follows that the effective potential can be approximated by the sum of the Casimir 
and the electrostatic potentials only when the last term in (\ref{omexap0}) is neglected. When the chemical nature of the 
anion and the cation is the same, this term is of a purely entropic origin.

When $\Delta {\cal L}[\vartheta_2,\phi]$ in Eq.~(\ref{omexap0}) is neglected, then the Casimir and the electrostatic 
terms are decoupled, and the concentration of the solvent and the solute density are obtained by the minimization of 
${\cal L}_C[\vartheta_1,\vartheta_2]$, whereas the charge profile is obtained by  the minimization of ${\cal L}_{DH}[\phi]$
 (with a simultaneous solution of the Poisson equation (\ref{Poisson})). 

When the last term in Eq.~(\ref{omexap0})  is taken into account,
 it directly yields an extra contribution to the effective
potential.  What is important, this term depends on both, $\kappa$ and $\xi$, as well as on the surface charges and 
the surface fields.
In addition, when $\Delta {\cal L}[\vartheta_2,\phi]$ is included, then the electrostatic contribution is 
 ${\cal L}_{DH}[\phi^{(1)}+\phi^{(2)}]$ rather than ${\cal L}_{DH}[\phi^{(1)}]$ obtained in the absence of the 
critical adsorption. This term also depends on $\xi$ and the surface fields through $\phi^{(2)}$ (see (\ref{eq:slit14})).
 We stress again that for a homogeneous solvent $\Delta {\cal L}[\vartheta_2,\phi]$ leads to the excess number density
 of ions in the layer of thickness $1/(2\kappa)$ \cite{israelachvili:10:0,barrat:03:0}. Neglecting this term 
leads to an oversimplified theory already for a homogeneous solvent. 

In this section we determine the form of the potential $\Psi(L)=\omega_{ex}-\gamma_0-\gamma_L$ by substituting  
to Eq.~(\ref{omexap0}) the
 approximate forms of the fields $\vartheta_i\approx\vartheta_i^{(1)}$ (Eq.~(\ref{vlin})) and 
$\phi\approx \phi^{(1)}+\phi^{(2)}$ (see Eqs.~(\ref{philin}) and (\ref{eq:slit14})). The expression for
 $\omega_{ex}[\vartheta_1,\vartheta_2,\phi]$ simplifies greatly when the fields $\vartheta_1,\vartheta_2,\phi$ 
satisfy the EL equations. For the Casimir part we obtain in our MF approximation
\begin{eqnarray}
\Psi_C={\cal  L}_C^0-\gamma_C(0)-\gamma_C(L)\approx A_C\xi^{-1}e^{-L/\xi},
\end{eqnarray}
where
$ A_C=-4H_i(0)J_{ij}H_j(L)
$ (see Eq.(\ref{Hi})).
 The remaining contribution to $\omega_{ex}$
 in the  approximation  consistent with Eq.~(\ref{omexap0}) can be written in the form (see (\ref{Lex}))
\begin{eqnarray}
\label{lex}
 {\cal L}_{el}= \Psi_{el}(L)+\gamma_{el}(0)+\gamma_{el}(L)\approx {\cal L}^{(1)}_{el}+ {\cal L}^{(2)}_{el}.
\end{eqnarray}
The leading order term ($O(\nu^2)$) is given by Eq.~(\ref{LDH}) with $\phi$ and $\psi$ approximated by the solutions
 $\phi^{(1)}$ and $\psi^{(1)}$ of the linearized equations,
\begin{eqnarray}
\label{el2}
{\cal L}^{(1)}_{el}=\int_0^L dz \Big[\frac{k_BT}{2\bar\rho_c}\phi^{(1)2} -
\frac{\bar\epsilon}{8\pi}\Big(\nabla \psi^{(1)}\Big)^2+e\phi^{(1)}\psi^{(1)}\Bigg]\\\nonumber
+e\sigma_0\big[\psi^{(1)}(0)+R_{\sigma}\psi^{(1)}(L)\big].
\end{eqnarray}
The well-known solutions are
\begin{eqnarray}
\beta \gamma_{el}^{(1)}(n)=\frac{\kappa\sigma^2_n}{2\bar\rho_c}
\end{eqnarray}
and
\begin{eqnarray}
\label{eq:DH}
\beta \Psi_{el}^{(1)}=\beta \Psi_{DH}=\frac{\kappa\sigma_0\sigma_L}{\bar\rho_c}\Bigg[\coth\Big(\frac{\kappa L}{2}\Big)-1\Bigg].
\end{eqnarray}

\vspace{1cm}

%%%%%%%%%%%%%%%%%%%%%%%%%%%%%%%%%%%%%%%%%%%%%%%%%%%%%%%%%%%%%%%%%%%%%%%%%%%%%%%%%%%%%%%%%%%%%%%%%%%%%%%%%%%%%%%%%%%%%%%%%%%%%%%%%%%%%%%%%%
\begin{figure}[htp]
\includegraphics[scale=0.38]{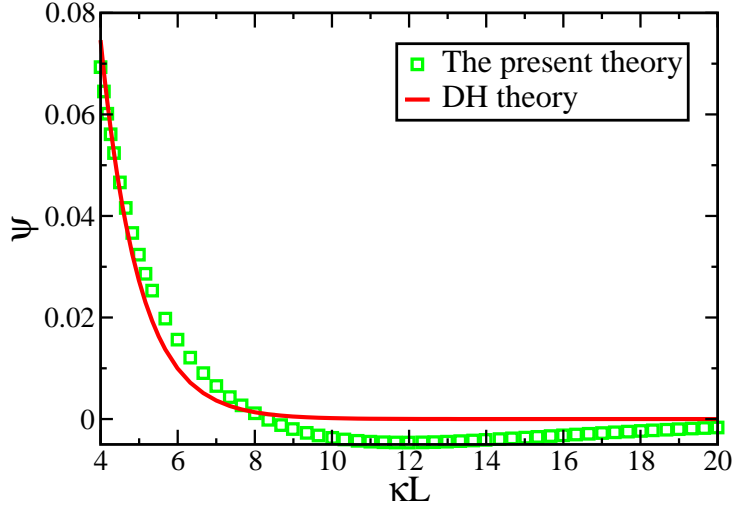}
\caption{The contribution to the effective  potential per unit area associated with the presence of charges, $\Psi=\Psi^{(1)}_{el}+\Psi^{(2)}_{el}$
(Eqs.~(\ref{eq:DH}) and (\ref{Lel3})) for $(-,-)$ BC as a function of the scaled distance $\kappa L$ (dotted line). 
Solid line is the potential $\Psi^{(1)}_{el}$ resulting from the linearized DH theory.
The potential is in  units
 of $\Big [\frac{\kappa \sigma_0^2 k_B T}{2\bar\rho_c}\Big ]$. $\kappa \xi=5$, $R_{\sigma}=R_n=1$ and $n_{01}=0.5$ 
(see Eqs.(\ref{eq:rsigma}) and (\ref{rn})).}
\label{fig:2}
\end{figure}
%%%%%%%%%%%%%%%%%%%%%%%%%%%%%%%%%%%%%%%%%%%%%%%%%%%%%%%%%%%%%%%%%%%%%%%%%%%%%%%%%%%%%%%%%%%%%%%%%%%%%%%%%%%%%%%%%%%%%%%%%%%%%%%%%%%%%%%%%%%

The leading-order correction term is of  the  order $O(\nu^3)$, and has the explicit form
\begin{eqnarray}
\label{el3}
 {\cal L}^{(2)}_{el}&=&\int_0^L dz \Bigg[-\frac{k_BT}{2\bar\rho_c^2}\phi^{(1)2}\vartheta_2^{(1)} +
\frac{k_BT}{\bar\rho_c}\phi^{(1)}\phi^{(2)} 
%\nonumber\\
%&-&
-\frac{\bar\epsilon}{4\pi}\nabla \psi^{(1)}\nabla \psi^{(2)}+e\big(\phi^{(1)}\psi^{(2)}+\phi^{(2)}\psi^{(1)}\big)\Bigg]
\nonumber \\
&+& e\sigma_0\big[\psi^{(2)}(0)+R_{\sigma}\psi^{(2)}(L)\big]
\end{eqnarray}
By using Eqs.~(\ref{psi1}) and (\ref{psi2}), integrating by parts and after some algebra we obtain 
\begin{eqnarray}
\label{Lel2}
{\cal L}^{(2)}_{el} &=& -\frac{k_BT}{2\bar\rho_c^2}\int_0^L dz \phi^{(1)2}\vartheta_2^{(1)}     \nonumber \\
&+&e\psi^{(2)}(L)\Bigg(R_{\sigma}\sigma_0+\frac{\nabla\phi^{(1)}(L)}{\kappa^2}
\Bigg)+e\psi^{(2)}(0)\Bigg(\sigma_0-\frac{\nabla\phi^{(1)}(0)}{\kappa^2}\Bigg),
\end{eqnarray}
where the first term equals $\Delta{\cal L}[\vartheta_2^{(1)}\phi^{(1)}]$.
The remaining terms come 
from ${\cal L}_{DH}[\vartheta_2^{(1)},\phi^{(1)}+\phi^{(2)}]$.
(The full expression for ${\cal L}^{(2)}_{el}$ is given in Appendix C.) \\
  We neglect terms $O(\exp(-2L/\xi),\exp(-2\kappa L))$, and after subtracting the surface-tension contributions, we
 obtain the approximation
\begin{eqnarray}
\label{Lel3}
\beta {\Psi}_{el}^{(2)} \approx -\frac{\kappa\sigma_0\sigma_Ln_{01}}{2\bar\rho_c^2}\Bigg\{  
A_1(y)e^{-L/\xi}
+A_2(y)e^{-\kappa L}+A_3(y)e^{-L/\xi}e^{-\kappa L}\Bigg\}
\end{eqnarray}
where the coefficients are (see Eqs.~(\ref{eq:rsigma}),  ~(\ref{rn}) and (\ref{y}))
\begin{eqnarray}
A_1(y)=\frac{R_{\sigma}^2+R_n}{R_{\sigma}}\Bigg(\frac{y}{2y-1}\Bigg)
\end{eqnarray}
\begin{eqnarray}
 A_2 (y)=(1+R_n)\Bigg(\frac{4y^2+4y}{2y+1}\Bigg)
\end{eqnarray}
\begin{eqnarray}
 A_3 (y)= -4(1+R_n)\Bigg(\frac{4y^3-2y}{4y^2-1}\Bigg).
\end{eqnarray}
The above approximation is valid for finite $y$; for $y\to\infty$ it is not justified to neglect terms  $O(\exp(-2L/\xi))$, 
and the approximation (\ref{omexap0}) is oversimplified. The case relevant for the experiments in Ref.~\cite{nature,nature_long}, 
however, corresponds to  $1<y\lesssim 10$. The effective potential $\Psi^{(1)}_{el}+\Psi^{(2)}_{el}$ (Eqs.(\ref{eq:DH}) and (\ref{Lel3})) is shown in
 Fig.~\ref{fig:2} for  $(-,-)$ BC. 
 A similar electrostatic attraction between likely charged surfaces was found in Ref.\cite{st:2011} in a nonlinear theory
 for large $\kappa\xi$.

 The final approximate expression for the potential per unit surface area of the slit, in the region  accessible in these experiments takes the form
\begin{eqnarray}
\label{Psifin}
\beta \Psi(L) \approx {\cal D}_1(y)e^{-L/\xi}+{\cal D}_2(y) e^{-\kappa L}+
{\cal D}_3(y) e^{-\kappa L}e^{-L/\xi}
\end{eqnarray}
where (see Eq.(\ref{rn}))
\begin{eqnarray}
\label{d1}
{\cal D}_1(y)=A_C\xi^{-1}-\frac{\kappa\sigma_0\sigma_Ln_{01}}{2\bar\rho_c^2}A_1(y)
\end{eqnarray}
\begin{eqnarray}
\label{d2}
 {\cal D}_2(y)=\frac{2\kappa\sigma_0\sigma_L}{\bar\rho_c}\Bigg[ 1-\frac{n_{01}}{4\bar\rho_c}A_2(y)\Bigg]
\end{eqnarray}
\begin{eqnarray}
\label{d3}
 {\cal D}_3(y)=-\frac{\kappa\sigma_0\sigma_Ln_{01}}{2\bar\rho_c^2}A_3(y).
\end{eqnarray}
In  Eqs.(\ref{d1})-(\ref{d3}) the excess number density of ions at one surface, $n_{01}=-H_2(0)$, and the number density 
of ions in the bulk, 
 $\bar\rho_c$, have the dimension of $1/volume$, surface number densities of elementary charges,
 $\sigma_0,\sigma_L$ have the dimension of $1/area$ 
(see Eq.(\ref{DH})), and the inverse Debye and correlation lengths, $\kappa$ and $1/\xi$ respectively,
 as well as the  amplitude
 $A_C$, have the dimension of $1/length$.

The above MF result can be corrected, because as  noted in  Sec.~\ref{subsec:sep} the Casimir contribution to the potential can be 
considered separately, and the critical fluctuations can be incorporated in this part. The universal  scaling function of 
the critical Casimir force has been obtained in Monte Carlo simulations \cite{vasiliev}. 
The amplitude $ A_C$ characterizing the long-distance decay of the potential has been extracted from the asymptotic behavior 
of this function for $L/\xi\gg 1$ in Ref.\cite{nature_long}. 
In the case of the  symmetrical BC ((+,+) or (-,-))   $A_C = A_+/\xi=-1.51(2)/\xi$,
 and  in the case of the antisymmetrical BC ((+,-) or (-,+))  $A_C= A_-/\xi=1.82(2)/\xi$ \cite{nature_long}.

 Important consequence of the coupling between the critical adsorption and charge distribution is the dependence of 
the prefactors in Eq.(\ref{Psifin})
on the ratio between the correlation and the screening lengths, $y$. 

\section{\bf Comparison with the experiment}
\label{sec:exp}

\subsection{Derjaguin approximation}
\label{subsec:derj}

\begin{figure}[t]
\centering
  \includegraphics[scale=0.4]{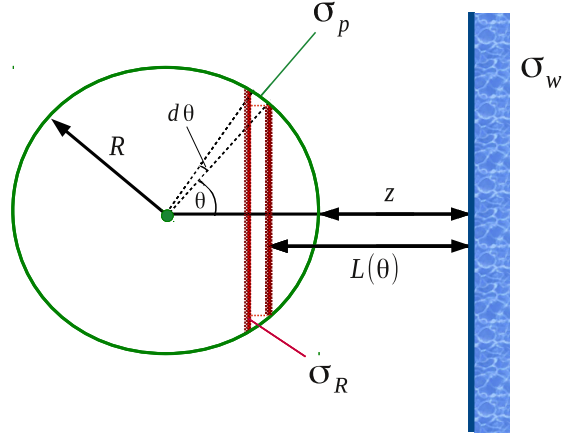}
\caption{Illustration of the  Derjaguin approximation for the plate-sphere geometry. $\sigma_P$, $\sigma_R$, $\sigma_w$ are the charge density of the
 spherical particle, of the ring, and of
 the wall respectively. $z$ is the minimal separation between the surface of the colloid and
the planar wall. $L(\theta)$ is the normal distance between the ring and the wall.
}
\label{Der}
 \end{figure}

The theory developed in the previous sections concerns confining surfaces that are planar and parallel to each other,
 whereas the measurements  in Refs.~\cite{nature,nature_long} with which we would like to compare our findings 
 were performed for a planar substrate and a spherical colloidal particle. 
When the colloidal particle radius is much larger than the separation of its surface from the substrate, then  the Derjaguin
 approximation can be applied, as in Refs.\cite{nature,nature_long}. 
The curved surface is approximated by a set of concentric circular rings of the  infinitesimal area $dS(\theta)$. 
The rings  are parallel to the substrate and are at the normal distance $L(\theta)=z+R(1-\cos\theta)$ 
(Fig.~{\ref{Der}). For each ring the excess grand potential per unit area is given in Eq.~(\ref{Psifin}), 
 except that the surface charge $\sigma_R$ of the ring differs from $\sigma_P$ of the colloidal particle, and the relation between them is 
\begin{equation}
\label{Derj2}
\sigma_{R}=\sigma_{P} / \cos\theta.
\end{equation}
Consequently,  the ratio between the surface charge at the substrate and at the ring is related to the corresponding ratio between the surface charge at 
the substrate and the particle by
\begin{equation}
\label{Derj3}
R_{\sigma [w/R]}= R_{\sigma [w/P]} \cdot   \cos \theta,
\end{equation}
where the symbols $w, P, R$  denote the  wall, the colloidal particle and the ring, respectively. 
The contribution of the ring to the potential between the substrate and the particle has the form
 \begin{eqnarray}
\label{Derj1}
 d\hat \Psi(z) =dS(\theta)  \Psi (L(\theta))
\end{eqnarray}
where $dS(\theta)$ is the area of the infinitesimal ring. 
Finally, the total potential $\hat \Psi(z)$ is obtained by summing all the contributions $d\hat \Psi(z)$ 
of the circular rings up to the maximal angle $\theta _M=\pi/2$,

\begin{equation}
\label{Derj4}
\hat \Psi(z) =\int_0^{\theta_{M}} dS(\theta)   \Psi (L(\theta))
\end{equation}
or 
\begin{equation}
\label{Derj5}
\hat \Psi(z) =\int_0^{\pi/2}  2\pi R^{2} \sin\theta  \cos\theta  \Psi  (z+R(1-\cos\theta)) d\theta 
\end{equation}

\subsection{Fitting}
\label{subsec:fit}

 In this section  we shall compare the predictions of our theory with the experiments 
reported in Ref.~\cite{nature,nature_long}.  In the experiment one surface was a charged surface of a particle, 
and the second surface was a flat, likely charged substrate chemically treated to achieve
a desired adsorption preference.

Although the  theory developed here is of the mean-field type, 
the Renormalization Group (RG) results can be applied to the Casimir part according to the discussion in  Sec.~\ref{subsec:sep}.
 We shall assume that the general form of the potential, Eq.~(\ref{Psifin}), is a fair approximation, except that the
 correlation length should have the correct temperature dependence, $\xi=\xi_0\tau^{-\nu}$, i.e., with $\nu$ taking the 
three-dimensional value $0.63$ of the Ising universality class.
Moreover, we shall assume that the 
Casimir amplitude $A_C$ is given by the proper universal form associated with the Ising universality class.

 We shall compare our predictions with the experiment 
for all four pairs of  the boundary conditions: $(+,+)$, $(-,-)$, $(+,-)$ and $(-,+)$, where $(+)$ denotes a hydrophobic and $(-)$ 
denotes a hydrophilic surface; the left  and the right symbol in the pair refer to the particle and the substrate respectively. Unfortunately, neither the charge density $\sigma_w\equiv \sigma_L$ nor the surface fields $H_i(L), i=1,2$
could be measured experimentally.
Two kinds of colloidal particles were used: a hydrophilic  
with the unknown charge density and the radius $R=1200 (nm) $, and a  hydrophobic 
with the radius $R=1850 (nm) $. 
The surface fields $H_i(0), i=1,2$ characterizing the colloidal particles  are also unknown. 
According to the experimental conditions we assume that the 
surface charge $\sigma_P\equiv \sigma_0$ and the surface fields for the colloidal particles  of the same type are fixed. 
We thus impose strict constraints
 on $H_2(0)$ and $\sigma_0$ to take on the same values   for $(+,+)$ and $(+,-)$ BC,
and the same values (but of course different than in the previous case) for $(-,-)$ and $(-,+)$ BC. 
The same constrains are imposed for the parameters that describe the flat surfaces, i.e., we
require that $\sigma_L=\sigma_0R_{\sigma}$ and $H_2(L)=H_2(0)R_n$
are the same  for  $(-,-)$ and $(+,-)$ BC, and likewise  the same for 
 $(+,+)$ and $(-,+)$ BC.

In experiments of  Refs.~\cite{nature,nature_long}, ions in the solution were present due to water dissociation 
 in a salt free water-lutidine mixture. For this mixture,  according to Ref.~\cite{gm:92} the density of (monovalent) ions is about 
$\bar\rho_c\simeq 1.08 \cdot 10^{-3}  mol/l$. We use this value, although
it appears to be  a rather rough estimate, and consider $\kappa$ as a fitting parameter.

We take into account that the wall-particle distance $z$ was determined in experiments
 up to $z_0=\pm 30(nm)$; we assume that for the specific boundary 
condition (the same series of measurements) the shift between the actual
 and measured distance is fixed; the shift can differ from one series of measurements to another. 
In the fittings, we have tried to keep the same value for $\xi_0$ for {\it all} sets of boundary conditions.
 The
 amplitudes $\hat A_C=2\pi A_{\pm} R$  have been taken from Ref.~\cite{nature_long}; $A_{+}$ and $A_-$ are 
the  amplitudes governing the asymptotic decay ($L/\xi \gg 1$, $\tau >0$) of the universal scaling functions of the critical Casimir force 
for symmetrical and antisymmetrical BC in a slit, respectively;
  for the Ising universality class in $d=3$, the Casimir scaling functions in a slit geometry  were obtained by  MC simulation method  in Ref.~\cite{vasiliev}.
 
%%%%%%%%%%%%%%%%%%%%%%%%%%%%%%%%%%%%%%%%%%%%%%%%%%%%%%%%%%%%%%%%%%%%%%%%%%%%%%%%%%%%%%%%%%%%%%%%%%%%%%%%%%%%%%%%%%%%%%%%%%%%%%%%%%%%%%%%%%%%%%%%%%%%%%%%%%%%%%%%%%%%%%%%%%%%%
\begin{figure}[t]
\centering
  \includegraphics[scale=0.4]{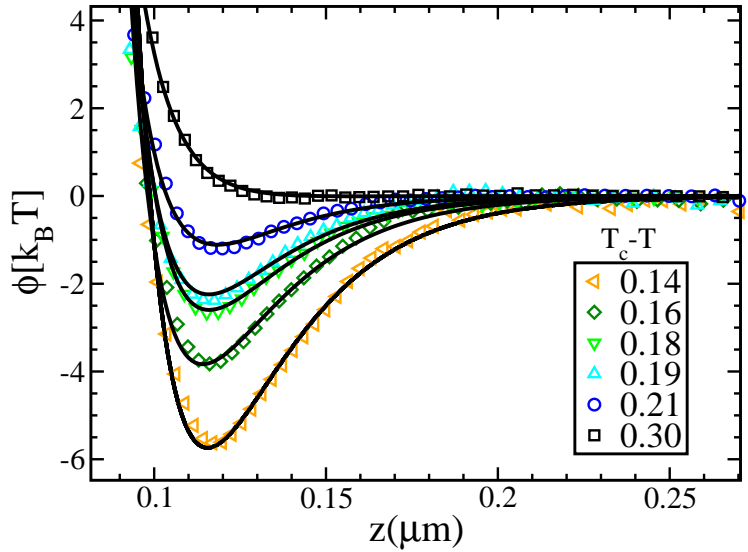}
\caption{(Color online) The effective potential between a  wall and a  colloidal particle of the radius  $R=1200 (nm)$ 
immersed in a water-lutidine mixture as 
 a function of  the distance $z$ for various temperatures $T$  \cite{nature_long}. 
The data refer to  $(-,-)$ BC corresponding to the case  where both  the colloidal particle and the substrate are hydrophilic. 
 Here $T_c$ is the critical temperature of the mixture. The solid lines are the
theoretical predictions (Eqs.~(\ref{Derj5})  and  (\ref{Psifin})) as  explained in the main text. The  parameters
 obtained from the fitting are shown in Table \ref{tab1}.
}
\label{fig:--}
 \end{figure}
%%%%%%%%%%%%%%%%%%%%%%%%%%%%%%%%%%%%%%%%%%%%%%%%%%%%%%%%%%%%%%%%%%%%%%%%%%%%%%%%%%%%%%%%%%%%%%%%%%%%%%%%%%%%%%%%%%%%%%%%%%%%%%%%%%%%%%%%%%%%%%%%%%%%%%%%%%%%%%%%%%%%%%%%%%%%%%%%%%%%%%%%%%%%%%%

In Fig.~\ref{fig:--}, we  show  the comparison between our theoretical predictions, 
 Eq.~(\ref{Derj5}) with $\Psi$ given by Eq.~(\ref{Psifin}) (solid lines),  and the experimental data  of Ref.~\cite{nature_long} 
for $(-,-)$ BC.
 The obtained fit parameters are given in Table \ref{tab1}. According to the table, 
the values of the correlation 
length 
  \begin{equation}
\label{Corr}
\xi=\xi^{fit}_0\left|\frac{T-T_c^{fit}}{T_c^{fit}}\right|^{-0.63}
\end{equation}
with $\xi_0^{fit}=0.21\pm 0.004 (nm)$ and   the Debye screening length $\kappa ^{-1}=10.9\pm0.6 (nm)$
  are both in the range of the experimental results. The best fit is obtained for a zero  shift in the critical
 temperature, $ T_c^{fit}=T_c $  but taking into account  up to $5 mK$ inaccuracy in $T$ itself.
The charge density of the colloid (in units of the elementary charge $e$)  obtained from the fit  is $\sigma_0=1.28 (nm)^{-2}$,
 which agrees nicely  with the value  given in  experiments of Refs.~\cite{gm:92,gm:92:1}, and is compatible with the observation that 
the highly charged colloids ($\gtrsim 0.24 (nm)^{-2}$) preferentially adsorb water while the colloids with  a smaller
 amount of charge prefer lutidine. 

%%%%%%%%%%%%%%%%%%%%%%%%%%%%%%%%%%%%%%%%%%%%%%%%%%%%%%%%%%%%%%%%%%%%%%%%%%%%%%%%%%%%%%%%%%%%%%%%%%%%%%%%%%%%%%%%%%%%%%%%%%%%%%%%%%%%%%%%%%%%%%%%%%%%%%%%%%%%
\begin{table}
\centering
\begin{tabular}{|l|c|c|c|c|c|c|c|c|c|}
\hline
 $T_c-T $ &$\xi (nm)$&  $\hat A_c(nm)$ &$\sigma_0 (nm)^{-2}$& $\sigma_L(nm)^{-2}$ &$ H_{2}(0)(nm)^{-3}$&$ H_2(L)(nm)^{-3}$ &$\kappa (nm)^{-1}  $&$z_0 (nm)$\\
\hline
0.14 &25.9& -11379&1.28&0.064&-0.01  &-0.0002 &0.0872&-30\\
\hline
0.16 & 24.3 &  -11379&1.28&0.064&-0.01  &-0.0002 &0.09&-30\\
\hline
0.18 &23.3&   -11379&1.28&0.064&-0.01  &-0.0002  &0.091&-30\\
\hline
0.19 &  22.85&   -11379&1.28&0.064&-0.01  &-0.0002&0.0918&-30\\
\hline
0.21 & 21.24&  -11379&1.28&0.064&-0.01  &-0.0002 &0.094&-30\\
\hline
0.30 &  16.3&  -11379&1.28&0.064&-0.01  &-0.0002  &0.097&-30\\
\hline
\end{tabular}
\caption{Fit parameters for the effective potential given by  Eqs.~(\ref{Derj5}) and (\ref{Psifin}) for $(-,-)$ BC
 where the colloidal particle
 with the  radius $R=1200 (nm)$ 
and the substrate are both hydrophilic. The amplitude $\hat A_C$ for this system is taken from Ref.\cite{nature_long}. 
 $\sigma_0$ and $\sigma_L$ denote the surface charge at the particle and at the substrate respectively in units 
of elementary charge $e$. $H_2(0)$ and $H_2(L)$ denote respectively the dimensionless effective potential per unit volume
 between the 
particle and ions, and the flat substrate and ions (see  Eqs.(\ref{Hi}) and (\ref{rn})).  $z_0$ is the experimental error in
the measured distance between the substrate and the particle. $\xi$ and $\kappa$ are the correlation and the inverse 
Debye-length 
respectively. See the main text for more details.}
\label{tab1}
\end{table}
%%%%%%%%%%%%%%%%%%%%%%%%%%%%%%%%%%%%%%%%%%%%%%%%%%%%%%%%%%%%%%%%%%%%%%%%%%%%%%%%%%%%%%%%%%%%%%%%%%%%%%%%%%%%%%%%%%%%%%%%%%%%%%%%%%%%%%%%%%%%%%%%%%%%
%

 Figure \ref{Fig(+,-)}  shows the experimental data (symbols)  and the theoretical curves (solid lines)
 for  the case of $(+,-)$ BC where the colloidal particle is 
hydrophobic whereas the wall is hydrophilic.  The obtained fit parameters are given in Table \ref{tab2}. According to the table, the 
amplitude of the correlation length 
and the Debye screening length are estimated as $\xi_0^{fit}=0.21\pm 0.001 (nm)$ and $\kappa ^{-1}=10 (nm)$, respectively. 
$H_2(L)$ and $\sigma_L$ are the same as for the $(-,-)$ BC.
In this case, the best fit is obtained with a shift in the critical
 temperature, $\Delta T_c^{fit}\equiv |T_c-T_c^{fit}|= 223 mK$ and  allowing up to $5 mK$ inaccuracy in $T$ itself. 
The charge density of the colloid (in units of $e$) obtained from fitting is $\sigma_0=0.7 \cdot 10^{-3} (nm)^{-2}$,
 which is in agreement with  Refs.~\cite{gm:92}  and  \cite{GHLB} reporting 
 the values for the surface charge densities of silica and polystyrene spheres in  water.  

%%%%%%%%%%%%%%%%%%%%%%%%%%%%%%%%%%%%%%%%%%%%%%%%%%%%%
\begin{figure}[t]
\centering
  \includegraphics[scale=0.4]{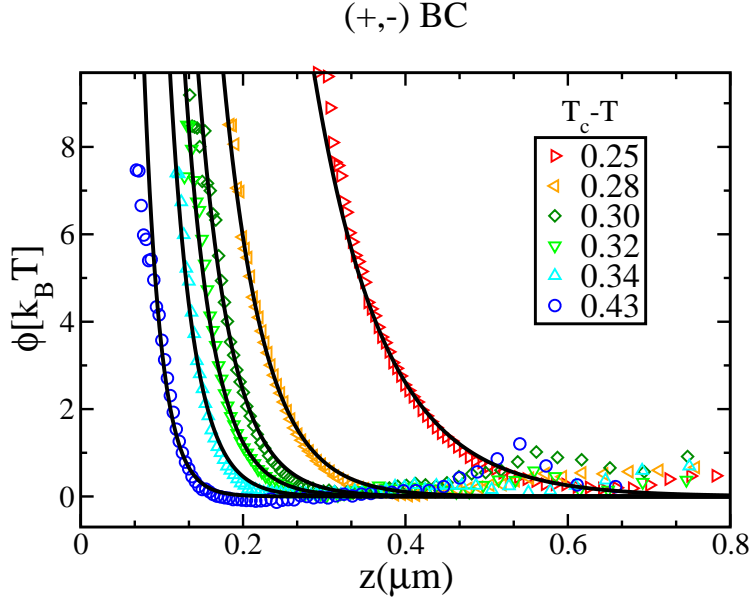}
\caption{(Color online) The same as in Fig.~\ref{fig:--} but for $ (+,-)$ BC 
corresponding to the hydrophobic colloidal particle and the hydrophilic wall. The radius of the colloidal particle is $R=1850(nm)$.
  The obtained parameters
 from the fitting are shown in Table \ref{tab2}.
}
\label{Fig(+,-)}
 \end{figure} 
%%%%%%%%%%%%%%%%%%%%%%%%%%%%%%%%%%%%%%%%%%%%%%%%%%%%%%%%%%%%
\begin{table}
\centering
\begin{tabular}{|l|c|c|c|c|c|c|c|c|c|}
\hline
 $T_c-T $ &$\xi (nm)$&  $\hat A_c(nm)$ &$\sigma_0 (nm)^{-2}$& $\sigma_L(nm)^{-2}$ &$ H_{2}(0)(nm)^{-3}$&$ H_2(L)(nm)^{-3}$ &$\kappa (nm)^{-1}  $&$z_0 (nm)$\\
\hline
0.25 & 85.77& 21144&0.0007&0.064 &0.001  &-0.0002 &0.1&-18\\
\hline
0.28 &49.5&  21144&0.0007&0.064 &0.001  &-0.0002&0.1&-18\\
\hline
0.30 &39.8&21144&0.0007&0.064 &0.001  &-0.0002&0.1&-18\\
\hline
0.32 &   34.8& 21144&0.0007&0.064 &0.001  &-0.0002&0.1&-18\\
\hline
0.34 & 29.4& 21144&0.0007&0.064 &0.001  &-0.0002&0.1&-18\\
\hline
0.43 &20.5& 21144&0.0007&0.064 &0.001  &-0.0002&0.1&-18\\
\hline
\end{tabular}
\caption{Fit parameters for the effective potential given in  Eqs.~(\ref{Derj5})  and (\ref{Psifin}) for $(+,-)$ BC
 corresponding to the 
hydrophobic  colloidal particle (with the radius $R=1850 (nm)$) and  
the  hydrophilic wall. The amplitude $\hat A_C$ for this system is  taken from Ref. \cite{nature_long}. For the 
remaining parameters see the caption of Table~\ref{tab1}.}
\label{tab2}
\end{table}

Similarly, for $(-,+)$ BC the comparison of the experimental data of \cite{nature_long} 
and our theoretical predictions for the effective potential are shown 
 in Fig.~\ref{Fig(-,+)}. The obtained fit parameters are given in Table \ref{tab3}. For this case, the estimates for the 
values of the amplitude of the correlation 
length 
and the Debye screening length  are  $\xi_0^{fit}=0.21\pm 0.003 (n m)$ and $\kappa ^{-1}=10 (nm)$, respectively. 
As for the $(-,-)$ BC, the best fit is obtained for considering no shift in  the critical
 temperature, $ T_c^{fit}=T_c $, and up to $5 mK$ inaccuracy in $T$ itself.
The charge densities and the surface fields of the colloidal particle and the substrate are consistent with  the results of fitting for 
$(-,-)$ BC.  

%%%%%%%%%%%%%%%%%%%%%%%%%%%%%%%%%%%%%%%%%%%%%%%%%%%%%%%%%%%%%%%%%%%
\begin{figure}[t]
\centering
  \includegraphics[scale=0.4]{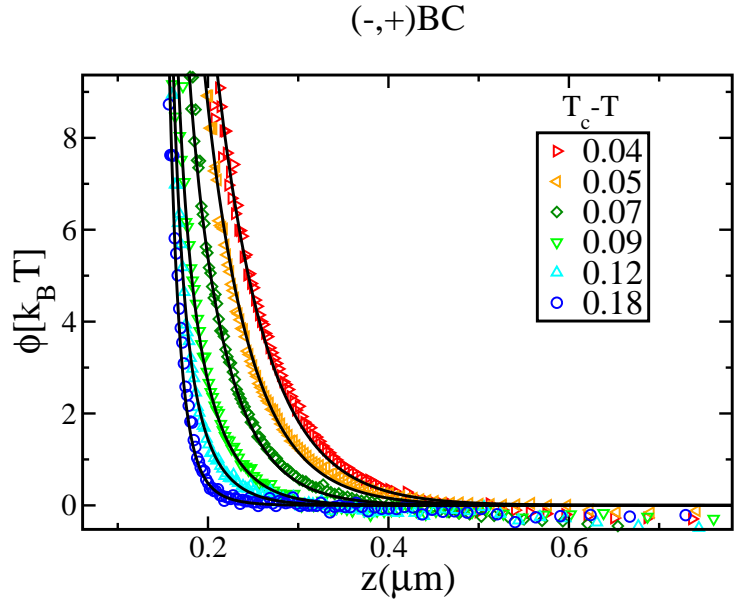}
\caption{(Color online) 
The same as in Fig.~\ref{fig:--} but for $ (-,+)$ BC 
corresponding to the hydrophilic colloidal  particle and the hydrophobic wall. The radius of the colloidal particle is $R=1200(nm)$.
}
\label{Fig(-,+)}
 \end{figure} 
%%%%%%%%%%%%%%%%%%%%%%%%%%%%%%%%%%%%%%%%%%%%%%%%%%%%%%%%%%%%%%%%%%%
\begin{table}
\centering
\begin{tabular}{|l|c|c|c|c|c|c|c|c|c|}
\hline
 $T_c-T $ &$\xi (nm)$&  $\hat A_c(nm)$ &$\sigma_0 (nm)^{-2}$& $\sigma_L(nm)^{-2}$ &$ H_{2}(0)(nm)^{-3}$&$ H_2(L)(nm)^{-3}$ &$\kappa (nm)^{-1}  $&$z_0 (nm)$\\
\hline
0.04 &55.07& 13715&1.28 &0.1344 &-0.01&0.0001 &0.1&30\\
\hline
0.05 & 50&13715&1.28 &0.1344 &-0.01&0.0001&0.1&30\\
\hline
0.07 & 42.5& 13715&1.28 &0.1344 &-0.01&0.0001&0.1&30\\
\hline
0.09 & 34.64&13715&1.28 &0.1344 &-0.01&0.0001&0.1&30\\
\hline
0.12 &29.64&13715&1.28 &0.1344 &-0.01&0.0001&0.1&30\\
\hline
0.18 &   23.50& 13715&1.28 &0.1344 &-0.01&0.0001&0.1&30\\
\hline
\end{tabular}
\caption{
Fit parameters for the effective potential given by  Eqs.~(\ref{Derj5}) and (\ref{Psifin}) for $(-,+)$ BC, where  the colloidal particle with
 radius $R=1200 (nm)$ 
is hydrophilic while the substrate is hydrophobic.
The amplitude $\hat A_C$ for this system is 
 taken from Ref. \cite{nature_long}. For more information see the caption of table I.
 }
\label{tab3}
\end{table}

For  the experimental data corresponding to the  $(+,+)$ BC, i.e., where both the colloidal particle and the wall are hydrophobic, 
we find that $1.1\leq\kappa \xi \leq 2.1 $ (see Table \ref{tab4}), which means that 
the approximation $\kappa \xi>>1$ under which we have obtained Eq.~(\ref{Psifin}) is not strictly valid. 
In this case, the terms which we have ignored in Eqs. (\ref{EL1a}) and  (\ref{EL2a}) (the terms which decay
 as $\exp(-2\kappa z)$) should 
be kept for  a better comparison with the experiment. 
Nevertheless, we have performed fitting  employing   the  relatively simple approximate form  (\ref{Psifin}) of an effective potential.
Because  the neglected  terms play  a more significant role for small distances, in Fig.~\ref{Fig(+,+)} 
 we report the comparison with  the experimental data only for distances larger than $z=80 (nm)$.  
 The obtained fit parameters are given in Table \ref{tab4}. According to the table
our estimate  for the amplitude of the correlation 
length is $\xi_0^{fit}\simeq 0.21\pm0.018 (n m)$ whereas for 
 the Debye screening length we have  $\kappa ^{-1}=14.35\pm 0.6 (nm)$. The smaller value obtained
 for $\kappa$ for this BC indicates,  according to Eq.~(\ref{kappa}),
 that the ion density $\bar \rho_c$  is smaller compared to the other BC. Since the experiment was performed for
 $\kappa\xi$ out of the  range of validity of our approximate result, further studies for $\kappa\xi\sim 1$ are required
 to verify whether the neglected terms in the potential would lead to the fit with $\kappa^{-1}\sim 10 nm$.  
For the present case, the best fit is obtained with a  shift in the critical
 temperature $\Delta T_c^{fit}\equiv |T_c-T_c^{fit}|= 63 mK$ and up to $15 mK$ inaccuracy in $T$ itself.
Again, the charge densities and surface fields of the colloidal particle and the substrate are consistent with the fitting parameters  obtained for  
$(-,+)$ and $(+,-)$ boundary conditions respectively.

%%%%%%%%%%%%%%%%%%%%%%%%%%%%%%%%%%%%%%%%%%%%%%%%%%%%%%%%%%%%%%%%%%%%%%%%%%%%%%%%%%%%%%%%%%%%%%%%%%%%%%%%%%%%%%%%%%%%%%%%%%%%%%%%%%%%%%%%
\begin{figure}[t]
\centering
  \includegraphics[scale=0.4]{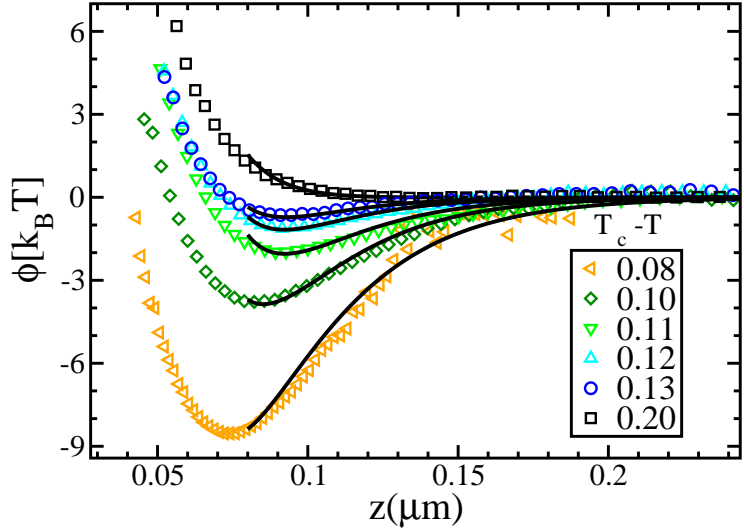}
\caption{(Color online)  The same as in Fig.~\ref{fig:--} but for $ (+,+)$ BC 
corresponding to hydrophobic both the  colloidal  particle and the  wall. The radius of the colloidal particle is $R=1850(nm)$.
The obtained parameters
 from the fitting are shown in Table \ref{tab4}.
}
\label{Fig(+,+)}
 \end{figure} 
\begin{table}
\centering
\begin{tabular}{|l|c|c|c|c|c|c|c|c|c|}
\hline
 $T_c-T $ &$\xi (nm)$&  $\hat A_c(nm)$ &$\sigma_0 (nm)^{-2}$& $\sigma_L(nm)^{-2}$ &$ H_{2}(0)(nm)^{-3}$&$ H_2(L)(nm)^{-3}$ &$\kappa (nm)^{-1}  $&$z_0 (nm)$\\
\hline
0.08 &30.1& -17543&0.0007 & 0.1344 &  0.001&0.0001    &0.07&-15\\
\hline
0.1 & 25.93& -17543&0.0007 & 0.1344 &  0.001&0.0001  &0.0675&-15\\
\hline
0.11 & 23.8&  -17543&0.0007 & 0.1344 &  0.001&0.0001  &0.0669&-15\\
\hline
0.12 & 21& -17543&0.0007 & 0.1344 &  0.001&0.0001&  0.069&-15\\
\hline
0.13 &19.3& -17543&0.0007 & 0.1344 &  0.001&0.0001  &0.0708&-15\\
\hline
0.2 &   15.8& -17543&0.0007 & 0.1344 &  0.001&0.0001 &0.0727&-15\\
\hline
\end{tabular}
\caption{Fit parameters for the effective potential given given by  Eqs.~(\ref{Derj5}) and (\ref{Psifin}) for $(+,+)$ BC, where the colloidal particle with 
radius $R=1850 (nm)$ and
 the substrate are both hydrophobic. Only the data  larger than  $z=80(nm)$ has been  considered. The amplitude $\hat A_C$ 
for this system is  taken from Ref. \cite{nature_long}. For the remaining parameters see the caption of Table I. }
\label{tab4}
\end{table}

%%%%%%%%%%%%%%%%%%%%%%%%%%%%%%%%%%%%%%%%%%%%%%%%%%%%%%%%%%%%%%%%%%%%%%%%%%%%%%%%%%%%%%%%%%%%%%%%%%%%%%%%%

We find that the best fit for $\xi_0$ for {\it all} BC is $\xi_0=0.21 (nm)$.  This value is in a very good agreement with 
several experimental results \cite{mb06,gcsp72,jbww87,elv93,slsl97}.
The correlation length  $\xi$ obtained from fits  given in Tabs~\ref{tab1}-\ref{tab4}  are compared with the 
expected behavior 
  \begin{equation}
\label{Corr2}
\xi=\xi_0\left|\frac{T-T_c}{T_c}\right|^{-0.63}
\end{equation}
 in Fig.~\ref{(-,-)XI}.

\vspace{2cm}
\begin{figure}[t]
\centering
  \includegraphics[scale=0.35]{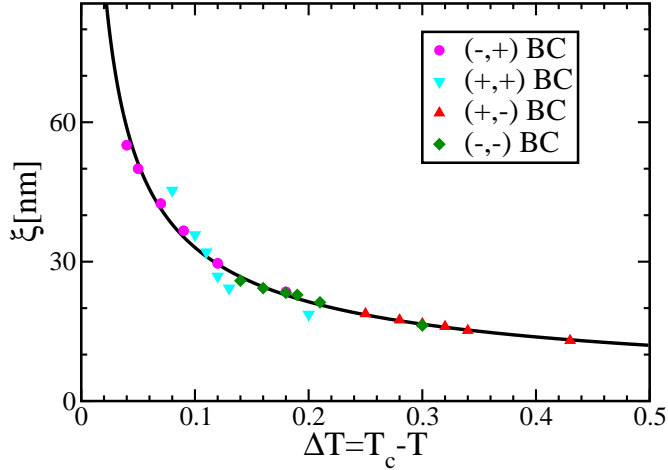}
\caption{The solid line shows the correlation length  $\xi$,   Eq.~(\ref{Corr2}) with the amplitude  $\xi_0=0.21 (nm)$
as a function of the deviation $\Delta T=T_c-T$ from the critical temperature $T_c$. 
Symbols 
correspond to the best fits  of the effective potential  given by  Eqs.~(\ref{Derj5}) and (\ref{Psifin})  (Table~\ref{tab1}-\ref{tab4}).
Recall that the case of (+,+) BC is out of range of validity of our approximation, and good agreement is not expected for this case.
}
\label{(-,-)XI}
 \end{figure}

 \section{\bf Summary and Discussion}
\label{sec:con}

In this paper we have studied   a  mutual effect of the critical adsorption and the distribution of ions on
 the effective potential between charged and selective surfaces confining a near-critical binary mixture with ions.
We have employed a Ginzburg-Landau-like theory  which  can be derived either from a lattice gas model ~\cite{cm:2010} or from a simple 
density functional theory~\cite{pc:2011} for a four component 
mixture. We assumed the same chemical nature of the anion and the cation, and a much bigger solubility of
 the ions in a one component of a binary solvent  than in the other.
Such conditions are met, e.g.,   for aqueous solutions in which  the ions come from a
dissociation of a water.  We have focused on the vicinity of the critical point of the solvent, where the correlation length
 for  fluctuations of the solvent concentration, $\xi$, is much bigger than the screening length $1/\kappa$. 

We have shown that when the chemical nature of the ions is the same, the excess grand potential  
of a system confined between two parallel walls can be split into two parts,  a
 part which is independent of the charge distribution, and a part which is independent of the solvent concentration. 
The first  contribution  describes a near-critical binary
solvent  with  a  neutral  solute (uncharged ions ) comprised of  single species  with a  preferential solubility in  water. 
This part yields  the critical Casimir
 potential. 
The second contribution to the excess grand potential is associated with the charge distribution, and has the form known from 
the DH theory. 
Both contributions, the critical Casimir and the DH,  depend on the number density of ions. 
 The equilibrium form of the number density of ions, corresponding to the 
minimum of the grand potential, is different from that which arises from the critical  Casimir part alone, and from the DH part alone. 
Thus, the effective potential between the confining walls differs from a sum of the Casimir potential in the uncharged 
system and 
the DH theory prediction for the electrostatic potential for ions in a homogeneous solvent.

We have shown that for $\kappa\xi\gg 1$  the effect of the critical adsorption on the charge distribution dominates, and the
 effect of charges on the solvent concentration can be neglected ~\cite{cm:2010}.
 This is because the 
screening length is much shorter than the correlation length, and the  charges that are present at
distances from the wall much smaller than $\xi$ can be neglected, like the other molecular details.
 Because of the preferential solubility in 
 water, the excess number density of ions decays in the same fashion as the excess solvent concentration, i.e. $\sim\exp(-z/\xi)$. 
The number density of ions determined by the critical adsorption is an input in calculations of the electrostatic
 contribution to the effective potential. 
 It influences the charge distribution, and in addition changes the entropic contribution in the DH part 
of the grand potential. As a result we obtain terms which were absent in the standard DH potential in the case of the homogeneous solvent. 
The dominant additional
 terms are $\sim\exp(-L/\xi)$ and $\sim\exp(-\kappa  L)$. 

Note that since the critical Casimir contribution to the excess grand potential 
has the same form as in the  uncharged critical system  and since
 for $\kappa\xi\gg 1$ it can be considered independently of the remaining contribution to the excess grand potential,
the RG theory can be applied to this part.
 Thanks to the above separation  and thanks to the analytical solution, 
we have been able  to  incorporate in our theory the RG results for the critical Casimir part. 
The remaining electrostatic part was obtained on the MF level, with {\it correct} 
form of the number density of ions resulting from the critical adsorption. 
The key result of the RG theory is the universal form of the critical Casimir potential, 
depending only on the boundary conditions as discussed in the Introduction. 
The molecular details, described in this theory by the vdW interaction potentials, 
influence only the nonuniversal properties, in particular the amplitude $\xi_0$ 
of the correlation function and the amplitude of the excess solvent concentration and the density of ions.

The dominant
  term in the electrostatic contribution to the excess grand potential decays in the same way as the critical Casimir potential.
This electrostatic contribution is nonuniversal. 
Even though the critical Casimir part of the excess grand potential should exhibit the universal behavior, 
it can be obtained from experiments only when the electrostatic contribution is subtracted. 

In order to verify the theory, we have fitted our predictions 
to the experimental results ~\cite{nature,nature_long}. 
Two different substrates and two different particles were used in experiments to yield 4 combinations of the boundary conditions.
 We have kept the same values of parameters characterizing the same surface.
 This requirement has provided us a constraint on the fitting parameters for the surface charge and  for the surface fields.
We have used the Derjaguin approximation to take into account the curvature of the  surface of the colloidal particle.
  We have obtained a good quantitative agreement  for a large range of distances (Figs.~\ref{fig:--}-\ref{Fig(-,+)})
 in three cases, and a less good  agreement (Fig.~\ref{Fig(+,+)}) in the fourth case, 
which is at the limits of applicability of the approximations we 
have made.  In our fitting,   the amplitude $\xi_0=0.21nm$ has the same value for all the considered cases, 
and this value is in a very good agreement with various experimental estimates ~\cite{mb06,gcsp72,jbww87,elv93,slsl97}. 
The best fit is obtained for the correlation length that agrees very well with the expected behavior (see Fig.~\ref{(-,-)XI})
(except from the boundary conditions $(+,+)$, where the agreement is less good).  
Also the fitted value of the critical temperature was precisely equal to the experimental value
 for $(-,-)$  and $(-,+)$ BC, with small shifts, $\Delta T_c=63mK$ and $\Delta T_c=223mK$,  for the $(+,+)$ and $(+,-)$ BC, respectively.

 As already mentioned in the Introduction, the attempt to fit experimental results  reported in Ref.~\cite{nature,nature_long}
to the sum of the critical  Casimir and the 
electrostatic potentials failed. 
Moreover, in order to fit the experimental results to the universal Casimir potential for large separations, 
a different value of  $\xi_0$ had to be chosen for each BC, from $0.17nm$ to $0.26nm$ for $(-,-)$ and $(-,+)$ BC respectively,
 and the 
 $(+,+)$ BC could not be fitted to the Casimir potential alone. 
A smaller value of $\xi_0$ leads to a larger prefactor $\hat A_C/\xi$ multiplying the exponential decay $exp(-L/\xi)$ of the 
potential. 
 In the present theory the prefactor of the  decay $exp(-L/\xi)$ contains the electrostatic contribution in addition to 
the universal 
Casimir amplitude. Therefore both, the present theory with $\xi_0=0.21nm$ and the  pure critical 
 Casimir potential with $\xi_0=0.17nm$ can yield  a  good fit at  large distances for  $(-,-)$ BC.
 However, unlike in the present approach, the other BC could not be fitted by the  critical Casimir potential alone
 with the same value of $\xi_0$, and the relative difference between the fitted amplitudes
 was as large as $(0.26-0.17)/0.17>50\%$.  
We cannot find  explanation for such large differences in the amplitudes for essentially the same mixtures. Boundary
 conditions
should not have any effect on the bulk properties. 
Moreover, in Ref.~\cite{nature,nature_long} only distances much larger from the position of the potential minimum could be
 fitted, whereas the present  theory
yields  a  good quantitative agreement for a wide range of distances. 
The agreement is obtained for all the measured systems for the parameters that agree very well with the experimental data,  and if precise experimental data are absent, are of a correct order of magnitude. There are no data for the surface fields of the four surfaces. 
The four free parameters, however, satisfy all the constraints of consistency for the four pairs of surfaces  (24 curves in 4 series of measurements).

We conclude by stressing that a very important advantage of the analytical expression  (Eqs.(\ref{Psifin})-(\ref{d3})) is the 
possibility of
designing the effective potential of a desired form by adjusting the surface charges and/or hydrophilicity
(or hydrophobicity) of the
surfaces, or the amount of ions in the solution.  Eq.(\ref{Psifin}) may  be a very useful tool in guiding 
future experimental studies.

\acknowledgments

We have greatly benefited from discussions with Siegfried Dietrich, Ursula Nellen, Marcus Bier and especially Laurent Helden, 
who explained us the details of experiments and commented on the results of fitting. Faezeh Pousaneh would like to thank 
Prof. Dietrich and his group for hospitality during her stay in Stuttgart, where a part of this work was done.
The work  of Faezeh Pousaneh  was realized within the International PhD Projects
Programme of the Foundation for Polish Science, cofinanced from
European Regional Development Fund within
Innovative Economy Operational Programme "Grants for innovation".

\section{Appendix A. Explicit expressions for $C^0_{ij}$}

The coefficients in Eq.~(\ref{C0}) take the explicit forms
\begin{eqnarray}
 C^0_{ss}= k_BT\frac{1-\bar\rho_c}{(1-\bar\rho_c)^2-\bar s^2}-6J_{ss}\\
C^0_{\rho\rho}= k_BT\Bigg(\frac{1-\bar\rho_c}{(1-\bar\rho_c)^2-\bar s^2}+\frac{1}{\bar\rho_c}\Bigg)-6J_{\rho\rho}\\
C⁰^0_{s\rho}=C^0_{\rho s}=k_BT\frac{\bar s}{(1-\bar\rho_c)^2-\bar s^2}-6J_{\rho s}.
 \end{eqnarray}
\section{Appendix B. Explicit expressions for the coefficients in Eq.~(\ref{eq:slit14}) }

The coefficients in Eq.~(\ref{eq:slit14}) take the explicit forms
\begin{eqnarray}
\label{eq:slit15}
{{\cal A}}_{0}=\left( \frac{{\cal A}_{00}}{y+1}\right)\left[ 1-y\frac{ (1-e^{-L/\xi})}{1-e^{-\kappa L}}e^{-\kappa L}\right]
-\left(\frac{{\cal A}_{0L}}{y-1} \right) \left[ 1-y\frac{1-e^{- L/\xi}}{1-e^{-\kappa L}}\right]
\end{eqnarray}
and
\begin{eqnarray}
\label{eq:slit16}
{{\cal A}}_{L}=\left( \frac{{\cal A}_{LL}}{y+1} \right) 
\left[ 1-y\frac{ (1-e^{-L/\xi})}{1-e^{-\kappa L}}e^{-\kappa L}\right]-\left(\frac{{\cal A}_{L0}}{y-1} \right)
 \left[ 1-y\frac{1-e^{-L/\xi}}{1-e^{-\kappa L}}\right]
\end{eqnarray}
where
\begin{eqnarray}
\label{eq:slit6}
 {\cal A}_{00}=n_0\frac{(y+1 )^2}{(2y+1)},
\end{eqnarray}
\begin{eqnarray}
\label{eq:slit7}
 {\cal A}_{0L}=-n_L\frac{(y-1)^2}{(2y-1 )}
\end{eqnarray}
\begin{eqnarray}
\label{eq:slit8}
  {\cal A}_{L0}=-n_0\frac{ R_{\sigma}(y-1 )^2}{(2y-1)}
\end{eqnarray}
\begin{eqnarray}
\label{eq:slit9}
  {\cal A}_{LL}=n_L
\frac{R_{\sigma}(y+1 )^2}{(2y+1)}
\end{eqnarray}

Note that each coefficient (\ref{eq:slit6})-(\ref{eq:slit9}) diverges for $y=\kappa\xi\to \infty$. However, the term
${\cal A}_{00}e^{-\kappa z}(1-e^{-z/\xi})$ in (\ref{eq:slit14}) remains finite, because when $\kappa\to\infty$ then $y e^{-\kappa z}\to 0$, 
and when $\xi\to \infty$ then $y(1-e^{-z/\xi})\simeq yz/\xi=\kappa z$. Likewise, the whole correction to the charge profile, 
Eq.~(\ref{eq:slit14}),  is finite.

\section{Appendix C. Explicit form of the leading-order correction term Eq.~(\ref{Lel2}) }

Full expression for the ${\cal L}^{(2)}_{el}$ in Eq.~(\ref{Lel2}) has following form
\begin{eqnarray}
 {\cal L}^{(2)}_{el}=\frac{k_B T \kappa \sigma_0^2}{2\bar \rho_c^2} \hat{\cal L}^{(2)}_{el}
\end{eqnarray}
where
\begin{eqnarray}
\hat{\cal L}^{(2)}_{el}=
\nonumber
\Bigg\{    \frac{2y(1-R_{\sigma})e^{-\kappa L}}{(1-e^{-\kappa L})^2}
\Bigg(\frac{(n_{0}-R_{\sigma}n_{L})(1-e^{-\kappa L}e^{-L/\xi})}{2y+1}-
\frac{(n_{L}-n_{0}R_{\sigma})(e^{-\kappa L}-e^{-L/\xi})}{2y-1}\Bigg)\\
\nonumber
- \frac{y(1-e^{-L/\xi})}{(1-e^{-\kappa L})^2}\Bigg(2R_{\sigma}(n_{0}+n_{L})e^{-\kappa L}+ \frac{(n_{0}+n_{L}R_{\sigma}^2)}{2y+1}e^{-2\kappa L}-
\frac{(n_{L}+n_{0}R_{\sigma}^2)}{2y-1}\Bigg)\\
\nonumber
+\Bigg(\frac{-2(n_{0}+n_{L})(1+R_{\sigma}^2)y^2+(n_{0}-n_{L})(1-R_{\sigma}^2)y}{4y^2-1}
\coth(\frac{\kappa L}{2})+\frac{y(n_{0}+n_{L}R_{\sigma}^2)}{2y+1} \Bigg)  \Bigg\}\\
\end{eqnarray}
and the expressions for $n_0$ and $n_L$ are given in Eq.~(\ref {n0}).

%%%%%%%%%%%%%%%%%%%%%%%%%%%%%%%%%%%%%%%%%%%%%%%%%%%%%%%%%%%%%%%%%%%%%%%%%%

\end{document}